\documentclass[10pt]{article}
\usepackage[dvips]{graphicx,color}

\newcommand{\ignore}[1]{}

\newcommand{\mComment}[1]{}
\newcommand{\dComment}[1]{}
\newcommand{\sComment}[1]{}
\newcommand{\rComment}[1]{}
\ignore{
\renewcommand{\mComment}[1]{\textcolor{blue}{Manny: #1}}
\renewcommand{\sComment}[1]{\textcolor{red}{Seth: #1}}
\renewcommand{\dComment}[1]{\textcolor{green}{David: #1}}
\renewcommand{\rComment}[1]{\textcolor{magenta}{Ray: #1}}
}

\newcommand{\ket}[1]{|{#1}\rangle}
\newcommand{\bra}[1]{\langle{#1}|}
\newcommand{\kets}[2]{|{#1}\rangle_{{}_{\!\!{#2}}}}
\newcommand{\bras}[2]{{}_{{}_{{#2}\!\!}}\langle{#1}|}
\newcommand{\slb}[2]{{#1}^{#2}}

\newcommand{\cT}{{\cal T}}

\newcommand{\trace}{\mbox{tr}}

\newcommand{\one}{{\mathchoice {\rm 1\mskip-4mu l} {\rm 1\mskip-4mu l} {\rm
1\mskip-4.5mu l} {\rm 1\mskip-5mu l}}}

\begin{document}

\title{
NMR Based Quantum Information Processing: \\
Achievements and Prospects
}
\author{
D.~G. Cory$^1$,  R. Laflamme$^2$, E. Knill$^2$, L. Viola$^4$, T.~F. Havel$^3$, N. Boulant$^1$, \\
G. Boutis$^1$, E. Fortunato$^1$, S. Lloyd$^4$, R. Martinez$^2$, C.
Negrevergne$^2$,\\ M. Pravia$^1$, Y. Sharf$^1$, G. Teklemariam$^5$, Y.~S.
Weinstein$^4$, W.~H. Zurek$^2$\\ 
\small$^1$Dept. of Nuclear Engineering, MIT, Cambridge, MA 02139, USA\\
\small$^2$Los Alamos National Laboratory, Los Alamos, NM, 87545, USA\\
\small$^3$BCMP, Harvard Medical School, 240 Longwood Avenue, Boston, Massachusetts
02115\\
\small$^4$Dept. of Mechanical Engineering, MIT, Cambridge, MA 02139, USA
\small$^5$Dept. of Physics, MIT, Cambridge, MA 02139, USA
}

\maketitle

\begin{abstract}
Nuclear magnetic resonance (NMR) provides an experimental
setting to explore physical implementations of quantum information
processing (QIP).  Here we introduce the basic background for understanding
applications of NMR to QIP and explain their current successes,
limitations and potential. NMR spectroscopy
is well known for its wealth of diverse coherent manipulations of spin
dynamics. Ideas and instrumentation from liquid state NMR spectroscopy
have been used to experiment with QIP.  This approach has carried
the field to a complexity of about 10 qubits, a small number for
quantum computation but large enough for observing and better
understanding the complexity of the quantum world.  While
liquid state NMR is the only present-day technology about to reach this
number of qubits, further increases in complexity will require new
methods.  We sketch one direction leading towards a scalable quantum computer
using spin 1/2 particles.  The next step of which is a solid state NMR-based
QIP capable of reaching 10-30 qubits.
\end{abstract}

\begin{flushleft}\begin{tabular}{l@{\ }l}
Corresponding author:&David G. Cory\\
&Massachusetts Institute of Technology\\
&NW14-2217\\
&150 Albany St\\
&Cambridge, MA 02139\\
&Phone: (617) 253-3806\\
&fax: (617) 253-5405\\
&\texttt{dcory@mit.edu}
\end{tabular}\end{flushleft}

\tableofcontents

\section{Introduction}

How can we turn quantum systems into useful and practical information processing devices?
At present we do not know a detailed implementation but we have taken important
steps both theoretically and experimentally to resolve this question.  The ultimate such device
will be a quantum computer (QC), and although such devices are  beyond today's experimental
reach, we do have related devices which are capable of performing information processing.

Quantum processors (QP)  use the laws of quantum mechanics for
computation, communication, and information
storage~\cite{bennett:qc1995a,divincenzo:qc1995b,divincenzo:qc1998b,lloyd:qc1995a,steane:qc1998b,bennett:qc1998b}.
What distinguishes quantum processors from classical ones
is the ability to operate quantum mechanically on superpositions of quantum states
and to exploit the resulting interference effects. To take advantage of QIP therefore
requires quantum mechanically controllable physical devices.  The
difficulty of building suitable devices cannot be overestimated: 
the extreme fragility of quantum information compared to its classical
counterpart makes robust manipulations much more difficult.
Furthermore, it is exactly these fragile states that needs to be harnessed
to achieve the potential of QIP. To turn QIP into a useful technology
requires exploring vast, new territories of quantum mechanics to
advance our understanding and our control of quantum mechanical
systems~\cite{divincenzo:qc2000a}. Fortunately, as this volume shows,
there is a rich variety of proposals for QIP implementations.

To evaluate, compare and measure the success of quantum devices for
QIP requires milestones on the long road towards building  a quantum 
computer.  A QP is a physical
device that operates via the laws of quantum mechanics, whose
evolution can be adequately controlled, with noise and
decoherence~\cite{zurek:qc1991a} held in check.  Ultimately this control will
lead to scalability, permitting fault tolerant computations such as Shor's
factoring algorithm or quantum physics simulations.  QPs can be compared on the
basis of their controllability, reliability, scale and efficiency.

While all proposals for QIP devices include provisions for reliable
control, to our knowledge scalability can only be achieved through
the key theoretical discovery of fault tolerant quantum computing.  
Before this discovery, it was suggested that
most computationally interesting
problems would never be solved using quantum
computers~\cite{landauer:qc1995a,unruh:qc1995a}: without methods
of error control quantum computers lose their power.  Fault tolerant
quantum computing is based on the existence of quantum error
correcting codes~\cite{shor:qc1995a,steane:qc1995a}. The main result is the accuracy threshold theorem~\cite{aharonov:qc1996a,kitaev:qc1997a,knill:qc1998a,preskill:qc1998a} which shows
that errors in control can be tolerated: it is possible to efficiently 
compute as accurately as desired provided the basic
device's precision exceeds a threshold.  Consequently, errors of real 
devices are not a fundamental
obstacle to QIP.  The accuracy threshold depends on the types of noise
that can affect a device and estimates for the maximum allowable error
probabilities range from $10^{-6}$ to $10^{-3}$. 
Fidelity of control is perhaps the most useful benchmark for comparing
QPs, being directly related to the scalability
of quantum devices.

In this contribution we discuss current and proposed roles for nuclear
magnetic resonance (NMR) based QIP, including liquid state, solid
state and engineered systems (see also the contribution by
Jones \cite{jones:qc2000b}).  The relationship between 
information theory and NMR extends back to the beginning of NMR's experimental explorations. In
1946, the groups of Purcell~\cite{purcell:qc1945a} and
Bloch~\cite{bloch:qc1946a} observed for the first time the magnetic
induction of nuclear spins.  This opened a new field of research
leading to many important applications such as molecular structure determination, dynamics studies both in the liquid and solid state~\cite{ernst:qc1994a},
and magnetic resonance imaging~\cite{mansfield:qc1982a}. Perhaps the
clearest example of early connections to information theory is the
spin echo~\cite{hahn:qc1950a} where Hahn demonstrated that
inhomogeneous interactions could be refocused to the extent that the 
phase of the nuclear spins retain information
about the local field.  This was followed by the magic
echo in the solid state~\cite{rhima:qc1970a} which inverts the dipolar
Hamiltonian so that the system can be returned to a prior state by
using the fact that the history of spin evolution is stored in
the density matrix.  Indeed, the idea of using nuclear spins for
information storage was suggested by Anderson and Hahn as early
as~1955~\cite{anderson:qc1955a,hahn:qc1955a}. 

Recently, NMR
was proposed as an accessible technology for
QIP~\cite{lloyd:qc1993a,cory:qc1996a,cory:qc1997b,chuang:qc1997a}.
The NMR proposals are unique in that the first generation devices
based on liquid state NMR  have already manipulated systems of up to seven qubits
using commercial technology.  The main achievement of
this technology has been to turn theoretical quantum algorithms into
experimental ones.  It is a technology where one can test the
assumptions of the theory and where one can learn to deal with the
imperfections of typical physical devices. The experiments have also
inspired new theoretical investigations of
QIP~\cite{knill:qc1997b,knill:qc1998c,knill:qc1999c,schulman:qc1998a,parker:qc2000a,ambainis:qc2000a}.  

Solid state NMR experiments provide a precedent for the controlled manipulation
of very high order quantum coherences.  For example, Pines and colleagues have
demonstrated that systems of hundreds of spin $1\over 2$ particles can be
coherently controlled~\cite{baum:qc1985a}.  In the solid state the
decoherence rate can be reduced through coherent averaging and phase
coherence times in the seconds have been achieved~\cite{cory:qc1990a}.
As the time for basic quantum operations is of the order of
100$\mu s$, the operational accuracy is in principle close to
what is needed for scalable fault tolerance.

One of the key lessons of NMR spectroscopy over the last four decades
and of NMR QIP over the last few years is that nuclear spin $1\over 2$ systems
are extremely robust quantum systems.  Their coherence can be
precisely manipulated and they decohere slowly on the time scale of
the qubit interactions.  Our present efforts towards building
a QP is based on nuclear spins, with key steps being:
\begin{itemize}
\item To develop liquid and solid state NMR methods for coherent control.
\item To move to solid state NMR, developing methods for achieving high
polarization.
\item To use engineered systems with the goal of increasing
clock rate and complexity.
\end{itemize}

This contribution is organized as follow. We introduce QIP, stressing the
fact that QIP devices, which are first and foremost controllable
quantum systems, emerge as an aspect of general principles of quantum
simulation.  The crucial elements required for understanding NMR QIP are
introduced.  These elements are dynamics, state initialization, and
readout.  We then explain the advantages and point out the limitations
of the current NMR QIP prototypes based on the liquid state.  To
overcome these limitations, we propose a next
generation of NMR QPs using solid state techniques.  This generation
will have very high polarization, faster gate times, slower
decoherence rates, and will include resettable qubits to
implement error correction.  We conclude by reviewing what
has been learned to date by NMR QIPs and give a brief discussion of
the required steps for constructing scalable quantum computers.

\section{Quantum simulation as a general approach to quantum information
processors}

In the first chapter of this volume, DiVincenzo describes the five 
requirements for a general purpose, scalable, quantum computer, they
can be summarized as follows:
\begin{itemize}
\item[\#1.] A scalable mapping of the basic units of QIP, qubits, into
physical systems.
\item[\#2.] The ability to initialize the state of the physical system to a known initial state
corresponding via the mapping to a standard initial state of the
qubits.
\item[\#3.] The availability of sufficient control via time-dependent
Hamiltonians for implementing a universal set of quantum gates.
\item[\#4.] The availability of sufficient control of the relevant noise mechanism, allowing the implementation of fault tolerant QIP. 
\item[\#5.] A qubit specific measurement capability.
\end{itemize}

\begin{figure}
\begin{center}
\includegraphics[height=5.0in,width=6.0in]{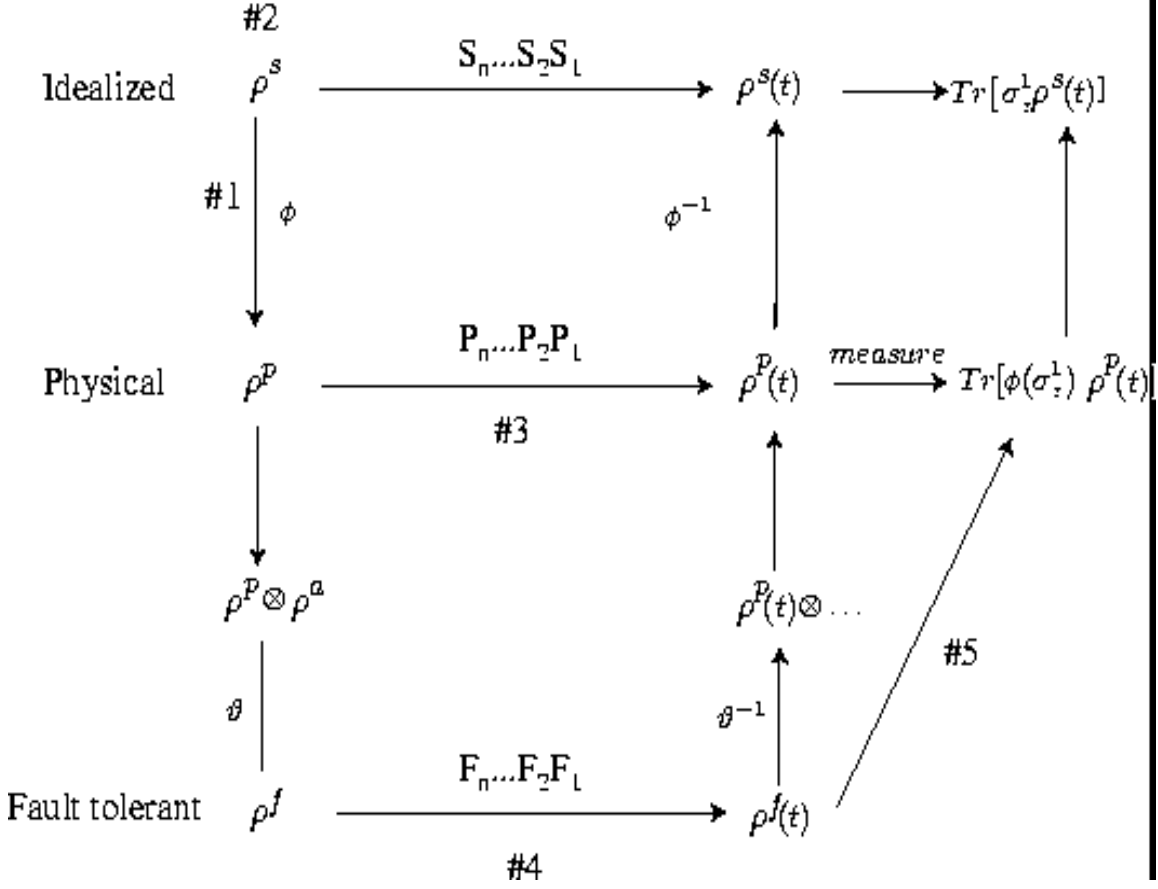}
\end{center}
\caption{Diagrammatic representation of quantum simulation.  The
system to be simulated ($S$), is mapped onto an ideal physical system
($P$) whose evolution is made to mimic that of the simulated system by
modulating the physical system's internal Hamiltonian using average
Hamiltonian theory. The mapping is linear and invertible, and any
Hamiltonian available in the physical space can be explored. The ideal
physical system is implemented by a real physical system ($P'$)
subject to noise.}
\label{figure:simphys}
\end{figure}

This description assumes that QC are based on qubits and states.  There is
a more general description using observables and subsystems which is especially
important for realizing fault tolerance in noisy quantum systems \cite{knill:qc1999b}.
Here we will conform to the state/qubit description but we believe that future QP will be based on an operator description, permitting the direct mapping of observables, and thereby the proper use of subsystems.
When designing QPs it is helpful to consider the five requirements in
the context of a general quantum simulation.  Quantum simulation is
best described by a diagram (Fig.~\ref{figure:simphys}) which relates
the abstract notion of a QP to its physical realization.  In this
diagram, the requirements correspond naturally to the crucial
components, as will be explained shortly.

The goal of QIP is to transform an initial state
$\rho^s$ by a series of quantum operations.  It is implicitly assumed that
the initial state is one that it is relatively easy to reach, such
as a thermal state or the ground state so that we do not hide the 
difficulty of reaching a final state in the starting state.
Once the final state is reached, we make a qubit specific measurement.    
We can restrict ourselves to measurements on a single qubit (e.g. qubit \#1) for
deterministic algorithms.  The process can be summarized by the following steps:
\begin{equation}
\rho^s \mathop{\longrightarrow}\limits^{S_m\dots S_2S_1}
\rho^s(t) \hfil \nonumber 
\mathop{\longrightarrow}\limits^{measure}
Tr\{\sigma_z^1\rho^s (t)\}.
\end{equation}

The task is to take a physical system and
engineer it in such a way as to fashion a QP from its
controlled dynamics.  Any QP is thus first and foremost a physical
system with its own internal and interaction Hamiltonians and modes of
decoherence. This physical system is used to in effect simulate
qubits, initial states, quantum gates, and measurements.  The diagram
of Fig.~\ref{figure:simphys} shows the relationship between the ideal
QP and the physical system in two steps (top to bottom), the first
involving an ideal, noiseless physical system, and the second showing
the process associated with the actual, noisy system.
The correspondence between DiVincenzo's five requirements and the key
steps in the simulation are described explicitly in the following subsections.

\subsection{Qubit}

The simplest building block for QIP is the
qubit \cite{schumacher:qc1996a}. A
qubit is a two level system with basic states $\ket{0}$ and $\ket{1}$
corresponding to the configurations of a classical bit. Qubits can be
represented by distinguishable spin $1\over 2$ particles to obtain a
straightforward mapping from the computational to the physical space.
When describing states of multi-qubits we will suppress
the tensor product operator between them so that 
$\sigma^1_z\otimes \sigma^2_z\otimes \one^3 $ will written as 
$\sigma^1_z\sigma^2_z\one^3$.

The states of the QP are related to states
of the (ideal) physical system  by the 
scalable mapping  $\phi$ (Fig.~\ref{figure:simphys}).
Scalability requires that the
mapping can be efficiently extended to cover additional qubits.

\subsection{Initial state}

The preparation of a known initial state
is accomplished by preparing the state mapped by $\phi$ in the
physical system,
\begin{eqnarray}
\rho^p &\mathop{\longrightarrow}\limits^{\phi^{-1}} \rho^s.
\end{eqnarray}

In most proposals for a QC requirement \#2, preparing a known starting
state, is achieved by cooling the system to the ground state either by
contact to a cold reservoir or by active means. Examples of cooling
methods in NMR include optical pumping \cite{navon:qc1996a} and
dynamic nuclear polarization \cite{wind:qc1985a,jeffries:qc1963a}.  An
alternative approach is to map the initial state of the system into a
state that behaves the same with respect to expectation value
measurements up to an overall scale factor. Such states are called
pseudo-pure states. They can be created from highly mixed
states states as was shown early in the development of NMR
QIP~\cite{stoll:qc1977a,cory:qc1997b,suter:qc1986a,chuang:qc1997a}.
Sec.~\ref{subsection:pps} gives a detailed implementation. Their use requires
sufficiently high signal-to-noise in the measurements which limits their scalability at
high temperature.  Nevertheless pseudo-pure states can be used to
investigate and benchmark small QPs, and this is the approach taken in
liquid state NMR.  On the other hand, some efficient algorithms
with no known efficient classical counterparts can be implemented
with initial highly mixed states which are scalable~\cite{knill:qc1998c}.

\subsection{Dynamical control}

In QIP we can think of applying control through a sequence of gates. The complete set of quantum gates are
constructed by using time-varying Hamiltonians in the physical system to
simulate the gates at each step. Thus the overall propagator in the ideal
physical system is $V_i=\phi(U_i)$.

A method to analyze coherent control using time-varying Hamiltonians has been  developed years
ago (much before the advent of QIP) by Waugh and
his colleagues~\cite{waugh:qc1982a,haeberlen:qc1968a,haeberlen:qc1976a} called
average Hamiltonian theory (AHT). AHT provides a quantitative approach to
understanding questions of coherent control and is complementary to
the gate based methods of QIP.  It enables one to solve for the evolution of the state
at a time $T$ by writing the evolution of a time independent average Hamiltonian $\bar H(T)$.
Average Hamiltonian theory separates the dynamics into 
a time invariant internal Hamiltonian, $H_{int}$, and 
a time dependent external Hamiltonian, $H_{ext}(t)$, giving
a total Hamiltonian of $H_{tot}(t) = H_{int}+H_{ext}(t)$.  The overall
dynamics after a period of evolution is then given by
\begin{equation}
U(T)= \cT e^{-i\int_0^{T} d\tau H_{tot}(\tau)} = e^{-i{\bar H}T},
\end{equation}
where $\cT$ is the Dyson time ordering operator.  The average
Hamiltonian defined by this equation is the time independent
Hamiltonian that would result in the same propagator if it were
applied over the same period. If the total Hamiltonian commutes with
itself at all times, $\bar H = {1\over T}\int_0^{T} d\tau H_{tot}(\tau)$.
This is rarely the case.  For sufficiently small $T$, the Magnus
expansion~\cite{magnus:qc1954a,wilcox:qc1967} provides a formal means of
calculating the average Hamiltonian:
\begin{equation}
\bar H =\bar H^{(0)} +\bar H^{(1)} + \bar H^{(2)} +\dots
\end{equation}
where,
\begin{eqnarray}
\bar H^{(0)} &=&{1\over T} \int_0^T  d\tau H_{tot} (\tau)\\
\bar H^{(1)} &=&{-i\over 2T} \int_0^T d\tau''\int_0^{\tau'}d\tau'[H_{tot} (\tau'),H_{tot}(\tau'')].
\end{eqnarray}
Examples of the use of AHT to implement quantum gates
are included in the following section.

The development of AHT is a prime example of how the development of NMR
anticipated many of the ideas of QIP using a different
language.  In the context of NMR Waugh demonstrated that:
\begin{itemize}
\item{} AHT can be used to design a control scheme for implementing
a desired average Hamiltonian,
\item{} Any average Hamiltonian (up to a scalar multiple)
can be implemented in most NMR systems of distinguishable
spins~\cite{glaser:qc1998a}.
\item{} The average Hamiltonian can be implemented to an arbitrary
precision if decoherence (random noise in the system and in the
control) is neglected.
\end{itemize}

AHT solves the problem of making available a complete set
of gates in the physical systems to meet requirement \#3.
The nuclear spins in a molecule can be chosen to be coupled
by internal Hamiltonians in a connected coupling network
(explained in Sec.~\ref{section:nmr}), and the results
of AHT given above show that in this case, a complete
set of gates can be implemented in principle.

There are many equivalent sets of gates that are complete for the
purpose of QIP. There are minimal sets consisting of one gate only,
but for efficient physical implementation, other sets are more
convenient~\cite{barenco:qc1996a,divincenzo:qc1995a,lloyd:qc1997a}.
The basic property of these gate sets are that any unitary
transformation can in principle be decomposed to arbitrary
precision into a sequence of gates from the set\footnote{Almost
all unitary transformations require exponentially many
gates to implement~\cite{barenco:qc1996a,knill:qc1995a}.}.
One such set consists of the operations on one and two qubits
given in the next two equations,
\begin{eqnarray}
U(\mathbf{v}\phi) &=& exp(-i(v_x\sigma_x+v_y\sigma_y+v_z\sigma_z)\phi/2),
\end{eqnarray}
where $\mathbf{v}=(v_x,v_y,v_z)$ is a real unit vector. This defines
a rotation by $\phi$ around $\mathbf{v}$ in the Bloch (or Poincar\'e) sphere
associated with the state space of a qubit.
The standard two qubit operation is the controlled-not defined by
\begin{equation}
U_{cnot12} = \ket{0}\bra{0}   \one^2+
\ket{1}\bra{1} \sigma_x^2.
\end{equation}
It flips the second qubit only if the first bit is in the state $\ket{1}$.
The gates in this set can be applied to any one or two qubits in the
computational space.  Example NMR implementations of these gates are
given in Sec.~\ref{section:nmr}.

The extent to which requirement \#3 is met in NMR depends on the
details of the system used. Since, as explained above,
the control is in principle sufficient to implement the desired
gates with arbitrarily accuracy, it is the extent to which
random noise affects the gates that is important. Probabilities
of errors per gate due to random noise can be as low as $.01$ in
liquid state systems for the most easily implementable two qubit
gates.

\subsection{Noise control} 

Noise can be controlled in principle through quantum error correction.
In order that fault tolerant method be effective it is sufficient
to be below an accuracy threshold, i.e. that the noise per gate
be smaller than a threshold value.   The threshold is  determined by the 
the types of errors and control we have over the device.  In order
to implement these fault tolerant procedures we have to increase
the size of the physical system but only by a ``reasonable'' amount
(polynomial in the size of th problem considered).

\subsection{Measurement}

 The measurement of one or more qubits translates via $\phi$ to
a corresponding measurement in the physical system.  Usually the
measurement is required to be projective. However, for most purposes,
it is sufficient to make a noisy expectation value measurement of a
single qubit (this works for all computations where a specific
answer is desired ~\cite{zhou:qc1999a}). Projective
measurements with known outcomes are in fact useful only for
improvements in efficiency.  Thus we require that it is possible to
measure $\trace[\slb{\sigma_z}{1}\rho^s(t)]$ with constant signal to
noise. If it were possible to measure this with infinite
signal to noise, i.e. exactly, then an apparently much more powerful
model of computation is obtained, with simple algorithms for solving
NP hard problems.  The quantity measured corresponds to
$\trace[\phi(\slb{\sigma_z}{1}) \rho^p(t)]$ in the
physical system.

The ability to make expectation value measurements as demanded
by our version of requirement \#5 is easily available in NMR
and is what makes NMR such a useful technology.

\section{Introduction to liquid state NMR}
\label{section:nmr}
 
NMR provides a valuable test bed for QIP ideas, since both the
internal Hamiltonian and the relaxation superoperator are well known
and reliable methods for controlling the dynamics are available.  No
other technology has so far reached the ability to control the number
of qubits accessible in NMR (currently seven).  
In addition, there are forty years of experience
in using NMR to explore chemical structure, dynamics and reactions.
Such experience has led to robust instrumentation capable of complex
experiments and to a wealth of practical methods for understanding
spin dynamics. 

\begin{figure}
\begin{center}
\mbox{
\includegraphics[height=4.0in,width=4.0in]{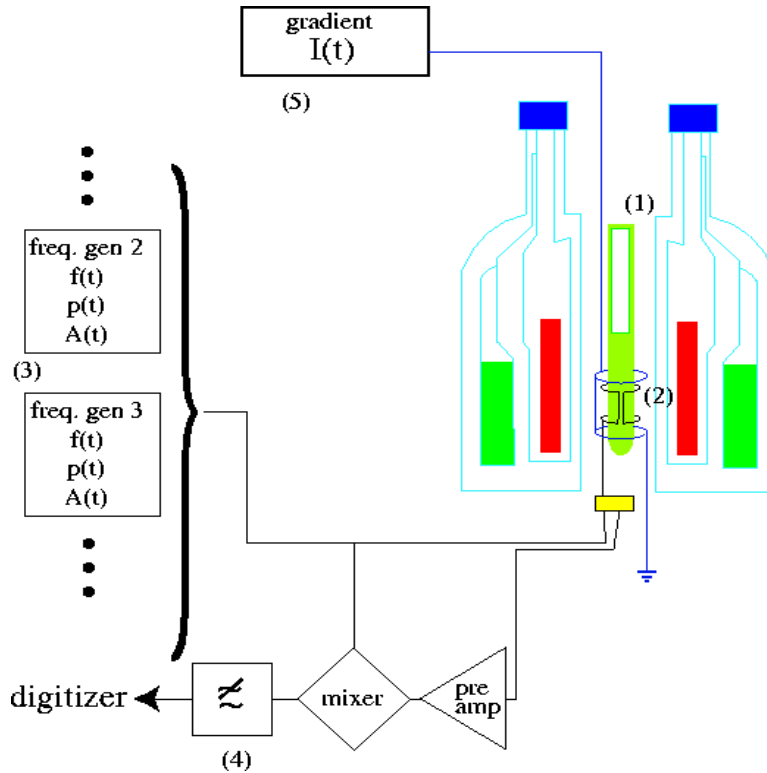}
}
\end{center}
\caption{A schematic block diagram of the key components of an NMR
system.  The sample (1) is placed in a strong, homogeneous magnetic field
(typically a superconducting magnet of 8--14 T) inside
a RF cavity (2) tuned to the resonance frequencies of
the nuclei of interest (typically $^1H$ and $^{13}$C).  At a field of
9.4 T, this corresponds to resonance
frequencies of 400 and 100 MHz respectively.  A parallel set
of computer-controlled RF transmitters (3) are connected to the cavity
with the frequency, phase, and amplitude of each channel independently
controllable.  The same RF cavity is used for detection, typically of
only one of the nuclear species ($^1H$ or $^{13}$C). The detected
signal (4) is referenced to the transmitter frequency, so only the
difference frequencies are observed.  Most modern NMR systems are also
provided with a programmable magnetic field gradient unit (5) which is a
convenient means for artificially introducing decoherence.  }
\label{figure:spec}
\end{figure}

A block diagram of an NMR spectrometer for QIP is shown in
Fig.~\ref{figure:spec}.  Even though the goals of standard NMR
experiments and QIP are quite different, the instrumentation
requirements are very similar and the same device can be used for both
purposes.  In liquid state NMR spectroscopy the goal is to
characterize the chemistry/dynamics of a solution containing known or
unknown molecules via a series of well characterized experiments
tailored to elicit specific components of the Hamiltonian or
relaxation superoperator in an easy to interpret fashion.  In NMR QIP
the goal is to characterize the effective overall transformation of a
complex series of operations on a very well characterized sample whose
internal Hamiltonian and relaxation superoperator are precisely known.
Ultimately the goal is to apply series of transformations sufficiently
complex  to be beyond the capacity of a classical computer.

\subsection{Qubits in NMR}
\label{subsection:nmrqubtis}

In NMR, the qubits are spin $1\over 2$ nuclei
in a molecule.  Although in principle other
nuclei can be used by observing that a spin $m$ nucleus can represent
$\log_2(2m+1)$ qubits, this however increases
the complexity of implementing the standard quantum gates given the
Hamiltonians that are usually available for control.

Distinguishable spin $1\over 2$ particles are convenient physical
qubits since, being two state systems, the mapping from the
computational space is usually straightforward and consists of associating
each qubit with one of the spins.  For the liquid and solid state NMR
devices the spins are distinguishable on the basis of the internal
Hamiltonian which is given by the chemistry of the sample.  For
engineered systems it is assumed that nuclear spins and thus qubits
can be spatially localized.

The robustness of spin $1\over 2$ particles is due to the fact that
they can interact directly only with magnetic, not with electric
fields.  This is particularly valuable since other than adjacent
qubits, sources of magnetic fields, will generally be distant and as a
result the interaction takes on a dipole-dipole symmetry.  Nuclear
spins have the additional advantage of being shielded by the electron
cloud and thus being very well isolated from most sources of
fluctuating magnetic fields, leading typically to long decoherence
times.

\subsection{Hamiltonians}
\label{subsection:hamiltonians}

NMR relies on a strong applied magnetic field to provide the dominant
terms of the internal Hamiltonian. The Zeeman interaction then
provides a convenient and uniform axis of quantization, with the sum
of the $\sigma_z$ giving useful quantum numbers. As a result, in the weak
coupling limit (i.e when the difference between chemical shifts is much larger than the coupling between spins), the eigenstructure of the internal Hamiltonian is easy
to determine.  The form of the internal Hamiltonian of a molecule's
nuclear spins is then to very good approximation,
\begin{equation}
H={1\over 2}\sum_i \omega_i\sigma^i_z +{\pi\over 2}\sum_{i\neq j} J_{ij}
\sigma^i_z\sigma^j_z,
\end{equation}
where the summation is over the nuclear spins.   
For liquid state NMR a typical sample contains $10^{18}$
molecules, each one acting as an independent QP.  Many copies are
necessary since the Zeeman interaction is very weak and approximately
$10^{15}$ spins are necessary to observe a signal.  The spectrometer
records the average state of the spins and the computation has to
preserve the Hilbert space spanned by a single molecule.  The through
space, spin-spin, dipolar interaction between molecules is averaged to
zero by random molecular motions on the time scale of the control and
measurement normally used, and is therefore only indirectly observable
via the effect on decoherence.

The form of the Hamiltonian is convenient for QIP.  The single
particle terms are naturally useful to distinguish qubits, while the
two-particle terms form the building blocks of conditional, two qubit gates. The
constants $\omega_i$ and $J_{ij}$ depend on the structure and electron
configuration of the molecule.  The resonance frequencies $\omega_i$
depend on the efficiency of screening of the nuclear spins from
the applied magnetic field by the surrounding electrons and the
scalar couplings $J_{ij}$ are mediated by electrons in molecular
orbitals that overlap both nuclear spins.

\begin{figure}
\begin{center}
\mbox{
\includegraphics[height=3.0in,width=5.0in]{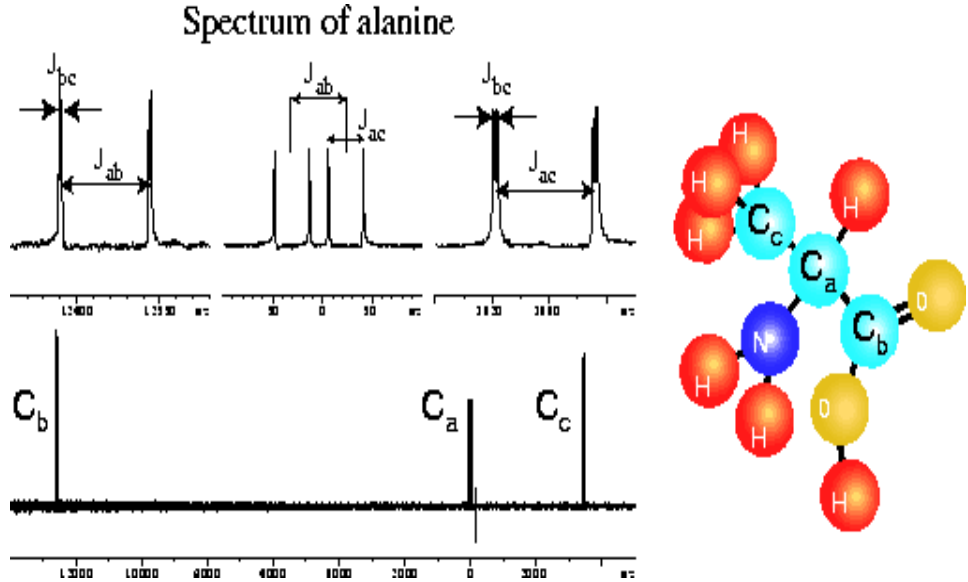}
}
\end{center}
\caption{$^{13}C$ proton-decoupled NMR spectrum of alanine at $9.4\textrm{T}$
showing the chemical shift differences between the three chemically
distinct carbon spins, and the couplings between them. The width of
the resonances provides an upper bound on the decoherence rates.  Note that
in the expanded views of the coupling between spins the horizontal scales
are not uniform.}
\label{figure:alanine}
\end{figure}

The constants of the internal Hamiltonian of the nuclear spins of a
molecule can be derived from an NMR spectrum acquired following a non-selective RF excitation pulse. Fig.~\ref{figure:alanine}
shows the spectrum for $^{13}C$ labeled alanine. (The $^1H$'s were
removed from the Hamiltonian (and spectrum) through spin decoupling,
described in Sec.~\ref{subsection:decoupling}.)  The constants can be
read off (in frequency units) from the positions of the peaks.  
Thus the three $^{13}C$ nuclei at $9.4\textrm{T}$ are
governed by the Hamiltonian
\begin{eqnarray}
H_{alanine} &=& \pi [10^8 (\sigma_z^a + \sigma_z^b + \sigma_z^c)
-12580\, \sigma_z^b + 3440\, \sigma_z^c] + {}\nonumber\\ 
&& (\pi/2)[53\,\sigma_z^a\sigma_z^b + 
38\, \sigma_z^a\sigma_z^c + 1.2\,\sigma_z^b\sigma_z^c],
\end{eqnarray}
where all of the frequencies are given in Hz. The last two terms
of the first line induce so-called chemical shifts which
distinguish qubits. The second line gives the
couplings between qubits.

The simplicity of the internal Hamiltonian also carries over to the
allowed modes of decoherence since there is no energy change
associated with a phase shift about the applied external field.  The
interaction of the environment thus only attenuates the off diagonal
elements of the density matrix (it does not mix these), and the
Liouvillian has a simple and sparse form. The relaxation
superoperator is simplest to describe in terms of a set of product
operators in the raising and lowering basis of the quantization axis.
The state of the system at any time on the other hand can be described as a sum of direct
products of the states of the individual spins, and these product
states have very simple transformation rules under both the internal
and external
Hamiltonians\cite{sorensen:qc1983a,vanderven:qc1983a,packer:qc1983a}.

In NMR the relaxation superoperator
\cite{redfield:qc1957a,redfield:qc1965a} is called the Redfield kite,
named after Alfred Redfield who first described it. It is called a kite
to provide a
descriptive name for the structure of the non-zero elements.  For a
three spin system, the superoperator has $64^2$ elements, of which
only 120 are non-zero in the absence of cross-correlation. The form of the relaxation superoperator
is summarized compactly in the following table.
\begin{center}
\begin{tabular}{l|ccccccccccccc}
& $\one$   & $\sigma_z$ &$\sigma_z$
& $\sigma_z$ &$\sigma_\pm$  &$\sigma_\pm$  
&$\sigma_\pm$  &$\sigma_\pm$  &$\sigma_\pm$  
&$\sigma_\pm$  &$\sigma_\pm$  &$\sigma_\pm$ 
&$\sigma_\pm$ \cr
& $\one$   &  $\one$ &$\sigma_z$
& $\sigma_z$ & $\one$  &$\sigma_z$  
&$\sigma_z$  &$\sigma_\pm$  &$\sigma_\pm$  
&$\sigma_\pm$  &$\sigma_\pm$  &$\sigma_\pm$ 
&$\sigma_\pm$ \cr
& \ $\one$   &  $\one$ &$\one$
& $\sigma_z$ & $\one$  &$\one$  
&$\sigma_z$  &$\one$  &$\sigma_z$  
&$\one$  &$\sigma_z$  &$\sigma_\pm$ 
&$\sigma_\pm$ \cr
\hline
$\one \one \one:1 $& X   &  X 
& X & X &  & 
& & & & & & & \cr
$\sigma_z \one \one:3 $& X   &  X 
& X & X &  & 
& & & & & & & \cr
$\sigma_z \sigma_z \one:3 $& X   &  X 
& X & X &  & 
& & & & & & & \cr
$\sigma_z \sigma_z \sigma_z:1 $& X   &  X
 & X & X &  & 
& & & & & & & \cr
$\sigma_\pm \one \one:6 $
& & & & & X & & & & & & & & \cr
$\sigma_\pm \sigma_z \one:12 $
& & & & & & X & & & & & & & \cr
$\sigma_\pm \sigma_z \sigma_z:6 $
& & & & & & & X & & & & & & \cr
$\sigma_\pm \sigma_\mp \one:6 $
& & & & & & & & X & & & & & \cr
$\sigma_\pm \sigma_\mp \sigma_z:6 $
& & & & & & & & & X & & & & \cr
$\sigma_\pm \sigma_\pm \one:6 $
& & & & & & & & & & X & & & \cr
$\sigma_\pm \sigma_\pm \sigma_z:6 $
& & & & & & & & & & & X & & \cr
$\sigma_\pm \sigma_\pm \sigma_\mp:6 $
& & & & & & & & & & & & X & \cr
$\sigma_\pm \sigma_\pm \sigma_\pm:2 $
& & & & & & & & & & & & & X \cr
& &  &  &  &  &  \cr
\end{tabular}
\end{center}
The $X$'s mark those elements of the relaxation superoperator that are
non-zero.  The number following the colon on the vertical axis
indicates the number of elements that are grouped together by
symmetry.  The grouping is convenient because if each spin sees the
same fluctuating field, then these elements will decohere at the same
rate.

Control over the spin state is achieved through externally
controllable Hamiltonians describing the interactions of the spin with
either RF pulses or magnetic field gradients.  The applied
Hamiltonian has the following form, assuming only one nuclear species:
\begin{equation}
H_{rf}= e^{i\omega_{trans}\sum_i\sigma^i_z/2} 
\Big ({\omega_1\over 2}\sum_i\sigma_x^i \Big )
e^{-i\omega_{trans}\sum_i\sigma^i_z/2}
\end{equation}
where $w_{trans}$ is the carrier frequency of the RF pulse and
$\omega_1$ is the effective amplitude of the RF field.
The RF pulses
can be applied non-selectively (to uniformly tilt all spins) by
choosing $\omega_1 \gg \Delta \omega$, or to selectively excite a
single spin by choosing and modulating $\omega_1$ subject to $\omega_1
\ll \Delta \omega$. Here $ \Delta \omega$ is the difference between
the transmitter frequency and that of the off-resonant spins.  The
selective version of $H_{rf}$ directly implements a useful version of
the single qubit gate, provided that the pulse time is kept short
compared to $1/J$, where $J$ is the largest coupling constant.

It is useful to perform computation with gates easily implemented by the
physical system.  For example, many quantum algorithms are written in terms of
Hadamard gates,
\begin{equation}
U_{Hadamard}={1\over \sqrt{2}}\left (\matrix{ 1 & 1\cr 1 & -1\cr}\right),
\end{equation}
which in terms of a time independent Hamiltonian corresponds to a $180^\circ$
rotation about the $\vec x + \vec z$ direction.  The Hadamard gate is used to rotate basis
states into superpositions.  Superpositions can also be reached through a
$90^\circ$ rotation about the $x$-axis,
\begin{equation}
U_{ x,90^\circ}={1\over\sqrt{2}}\left (\matrix{ 1 & -i\cr -i & 1\cr}\right),
\end{equation}
which is simpler to implement in NMR.  Often a computation can be
written in terms of gates that correspond more directly to the
physical system.

A second example of physical gates are selective z-rotations which may
be obtained from a series of RF pulses.  This uses the fact that rotations
around the $x$ and $y$ do not commute giving rise to z-rotations.  Another
possibility to implement z-rotation is to make a change of reference frame.
Referring back to the block
diagram of the spectrometer (and recalling that z-rotations commute
with the internal Hamiltonian) we see that z-rotations can be
accounted for in the reference phase of the individual transmitters
and thus in general no explicit pulses are required.  Keeping track of
z-rotations becomes a book-keeping task.

A somewhat neglected part of QIP is the ability to introduce selective
decoherence at will.  This is convenient for preparing specific states
and required for removing undesired information.  Since NMR methods
are ensemble measurements we may take advantage of the spatial
distribution of the sample to create qubit selective decoherence.
Decoherence is obtained by the action of a magnetic field gradient
followed by molecular diffusion to remove all recoverable spatial
correlation.  The essential feature is that the spectrometer is
normally equipped with a magnetic field gradient coil connected to a
programmable current source, so the Hamiltonian is,
\begin{equation}
H_{grad}= {\partial B_z(t) \over \partial z} z \sum {\gamma_j\over 2} \sigma^j_z,
\end{equation}
where $\gamma$ is the gyromagnetic ratio, and $z$ is the spatial
location of the spin along the gradient direction.  The action of the
gradient is to create a linear phase ramp throughout the sample
\cite{sodickson:qc1998a} which provides a decoherence mechanism with
relaxation rate given by
\begin{equation}
R= \gamma^2\left ({dB_z \over dz}\right )^2 \delta^2,
\end{equation}
where $\delta$ is the time during which the spin system evolved in the
magnetic field gradient, and the experiment consists of two strong,
short and (effectively) opposite gradient pulses separated by the
decoherence interval.

\subsection{Unitary gates}
\label{subsection:unitarygates}

We will now show that through a combination of the internal Hamiltonian,
modulated via time-dependent external Hamiltonians designed following
the prescriptions of average Hamiltonian theory, any desired unitary
operator can be achieved.  It is only necessary to do this for a
complete set of gates such as the C-not and single qubit gates
(described above), but we will start by exploring a no-operation gate.
A useful benchmark of QIP is the ability to suppress the internal
Hamiltonian and thus to leave the system in whatever state it started
in.  This will also permit us to describe coherent averaging in some
detail.

\subsubsection{No-operation}
\label{subsubsection:noop}
\label{subsection:decoupling}

\begin{figure}
\begin{center}
\mbox{
\includegraphics[height=1.5in,width=5.0in]{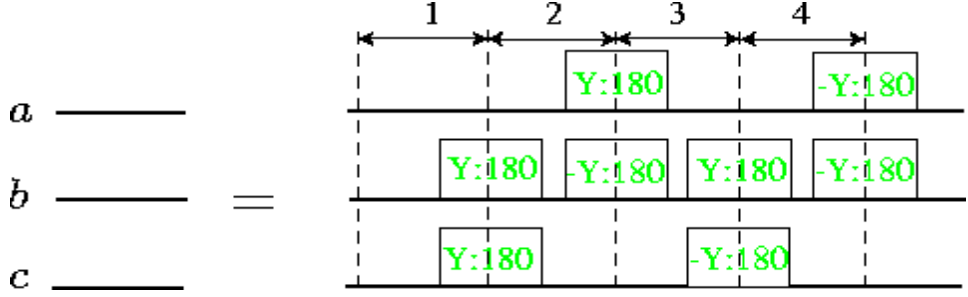}
}
\end{center}
\caption{
Implementation of a no-operation gate for three spins. The right-hand
side shows the sequence of RF pulses which accomplishes this
regardless of the Larmor frequencies and coupling constants.
The pulses are a series of equally spaced parallel 
$180^\circ$ rotations about the
$y$ axis. The net effect of the eight RF pulses is that except
for the effects of decoherence, the average propagator is the identity.
The left-hand side shows the equivalent computational gates as a staff
with each qubit identified}
\label{figure:noop}
\end{figure}

One interesting feature of NMR compared to a device like the ion trap is that the coupling between qubits is always active.  Refocusing
methods have been invented to selectively turn off undesired couplings,
so as to suppress their contribution to the effective propagator
over some time interval. For example, starting with 
the internal Hamiltonian for a three qubit system:
\begin{eqnarray}
H_{3\ qubits}={1\over 2}(\omega_a \slb{\sigma_z}{a} +\omega_b \slb{\sigma_z}{b}
+\omega_c\slb{\sigma_z}{c}) + \\
{\pi\over 2} J_{ab} \slb{\sigma_z}{a} \slb{\sigma_z}{b} + 
{\pi\over 2} J_{ac} \slb{\sigma_z}{a} \slb{\sigma_z}{c} +
{\pi\over 2} J_{bc} \slb{\sigma_z}{b} \slb{\sigma_z}{c} 
\end{eqnarray}
we can implement the identity operation 
using the sequence of RF pulses shown in Fig.~\ref{figure:noop}.

The effect of the sequence of Fig.~\ref{figure:noop} can be appreciated by
breaking up the evolution into its four equal time parts, assuming that
the RF pulses are effectively instantaneous, and transforming to a
time-dependent interaction frame defined by the RF pulses (termed toggling
frame Hamiltonian by Waugh~\cite{haeberlen:qc1968a}). The
interaction-frame form of the internal Hamiltonian can be broken up into
commuting parts. This can be visualized using the following table.  The
periods correspond to those shown in Fig.4 and the toggling frame states
(with tilde) correspond to the interaction frame states under the actions
of the RF pulses which are assumed to be delta function pulses.

\begin{center}\begin{tabular}{c|cccc|c}
term &  1   &  2 & 3 & 4 & average
\cr
\hline
$\tilde\slb{\sigma_z}{a}$ &  +$\slb{\sigma_z}{a}$   &  +$\slb{\sigma_z}{a}$
  & -$\slb{\sigma_z}{a}$& -$\slb{\sigma_z}{a}$& 0\cr
$\tilde\slb{\sigma_z}{b}$ &  +$\slb{\sigma_z}{b}$   &  -$\slb{\sigma_z}{b}$
& +$\slb{\sigma_z}{b}$& -$\slb{\sigma_z}{b}$ & 0 \cr
$\tilde\slb{\sigma_z}{c}$ &  +$\slb{\sigma_z}{c}$   &  -$\slb{\sigma_z}{c}$
& -$\slb{\sigma_z}{c}$& +$\slb{\sigma_z}{c}$& 0 \cr
$\tilde\slb{\sigma_z}{a}\tilde\slb{\sigma_z}{b}$ &  +$\slb{\sigma_z}{a}\slb{\sigma_z}{b}$ 
  &  -$\slb{\sigma_z}{a}\slb{\sigma_z}{b}$  & -$\slb{\sigma_z}{a}\slb{\sigma_z}{b}$ 
& +$\slb{\sigma_z}{a}\slb{\sigma_z}{b}$  & 0\cr 
$\tilde\slb{\sigma_z}{a}\tilde\slb{\sigma_z}{c}$ &  +$\slb{\sigma_z}{a}\slb{\sigma_z}{c}$   &
 -$\slb{\sigma_z}{a}\slb{\sigma_z}{c}$ & +$\slb{\sigma_z}{a}\slb{\sigma_z}{c}$
& -$\slb{\sigma_z}{a}\slb{\sigma_z}{c}$ & 0\cr 
$\tilde\slb{\sigma_z}{b}\tilde\slb{\sigma_z}{c}$ &  +$\slb{\sigma_z}{b}\slb{\sigma_z}{c}$   &
 +$\slb{\sigma_z}{b}\slb{\sigma_z}{c}$ & -$\slb{\sigma_z}{b}\slb{\sigma_z}{c}$
& -$\slb{\sigma_z}{b}\slb{\sigma_z}{c}$ & 0\cr 
\end{tabular}\end{center}

The terms of the Hamiltonian in the toggling frame Hamiltonian  are
shown for each of the four intervals.  The average Hamiltonian is
simply the time average of these states, since in this case the
various toggling frame Hamiltonians commute with each other over all periods.  The
action of the final pulses is not included in the table, but is
important because in the absence of the internal Hamiltonian the
effective propagator for the external Hamiltonian should be the
identity operator.

\begin{figure}
\begin{center}
\mbox{
\includegraphics[height=1.4in,width=4.2in]{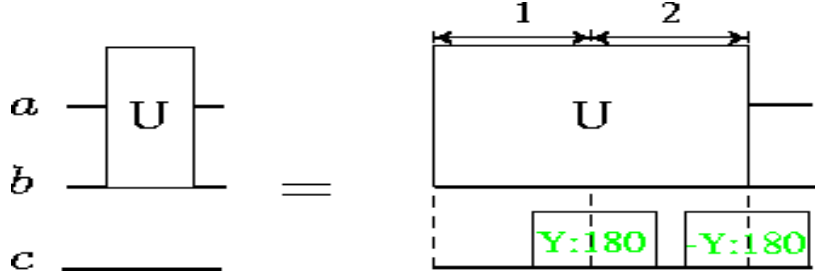}
}
\end{center}
\caption{Sequence of RF pulses which accomplishes decoupling of spin $c$ from 
spins $a$ and $b$. This effectively removes the $c$ qubit (or spin)
from the evolution.  Qubit $c$ therefore stays in its original state
while $a$ and $b$ evolve under their internal Hamiltonian.}
\label{figure:decouplingc}
\end{figure}

Similar techniques can be used to selectively ``turn off''
interactions to a subset of the nuclei. One example of
this can be used to simplify observation, by removing
couplings during observation. This decoupling technique
is widely used in NMR and can be thought of as an experimental
method for ``tracing out'' unwanted degrees of freedom,
as was done to obtain the spectra of Fig.~\ref{figure:alanine}.
In Fig.~\ref{figure:decouplingc}, a sequence of pulses demonstrate
how this can be done. The following table gives the transformations
of product operators of the 3 qubit Hamiltonian and their time
average.

\begin{center}
\begin{tabular}{c|cc|c}
term &  1   &  2 &  average
\cr
\hline
$\tilde\slb{\sigma_z}{a}$ &  +$\slb{\sigma_z}{a}$   &  +$\slb{\sigma_z}{a}$
& $\slb{\sigma_z}{a}$\cr
$\tilde\slb{\sigma_z}{b}$ &  +$\slb{\sigma_z}{b}$   &  +$\slb{\sigma_z}{b}$
& $\slb{\sigma_z}{b}$ \cr
$\tilde\slb{\sigma_z}{c}$ &  +$\slb{\sigma_z}{c}$   &  -$\slb{\sigma_z}{c}$
& 0 \cr
$\tilde\slb{\sigma_z}{a}\tilde\slb{\sigma_z}{b}$ &  +$\slb{\sigma_z}{a}\slb{\sigma_z}{b}$ 
  &  +$\slb{\sigma_z}{a}\slb{\sigma_z}{b}$    & $\slb{\sigma_z}{a}\slb{\sigma_z}{b}$\cr 
$\tilde\slb{\sigma_z}{a}\tilde\slb{\sigma_z}{c}$ &  +$\slb{\sigma_z}{a}\slb{\sigma_z}{c}$   &
 -$\slb{\sigma_z}{a}\slb{\sigma_z}{c}$  & 0\cr 
$\tilde\slb{\sigma_z}{b}\tilde\slb{\sigma_z}{c}$ &  +$\slb{\sigma_z}{b}\slb{\sigma_z}{c}$   &
 -$\slb{\sigma_z}{b}\slb{\sigma_z}{c}$  & 0\cr 
\end{tabular}\end{center}
The corresponding 3-qubit Hamiltonian is averaged to:
\begin{equation}
H_{3\ qubits}={1\over 2}
(\omega_a \slb{\sigma_z}{a} +\omega_b \slb{\sigma_z}{b}) +
{\pi\over 2} J_{ab} \slb{\sigma_z}{a} \slb{\sigma_z}{b}.
\end{equation}
Thus if the evolution of the system is followed under the action of
the average Hamiltonian one of the qubits appears to vanish from the
system when observing the spectral lines associated with
spins $a$ and $b$. For observation purposes it is sufficient to irradiate at
$c$'s resonance frequency with an RF field stronger than any of the
scalar couplings involving $c$.
Such decoupling techniques have inspired coherent approaches to other control
problems such as dynamical decoupling or error correction
for general noise~\cite{viola:qc1998a,viola:qc1999a,knill:qc1999b}.

\subsubsection{Control-not}
\label{subsubsection:cnot}

\begin{figure}
\begin{center}
\mbox{
\includegraphics[height=1.5in,width=4.2in]{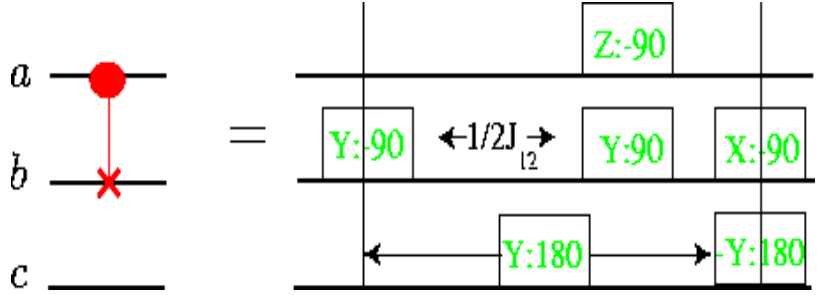}
}
\end{center}
\caption{Translation of a C-not from qubit $a$ to qubit $b$ into a pulse
sequence for nuclear spins. Note the required refocusing to suppress
evolution involving spin $c$}
\label{figure:cnot}
\end{figure}

The C-not gate is a rotation of one qubit conditional on the
state of a second.  The desired propagator thus has the simple form
\begin{eqnarray}
U_{cnot}&=&\left(\matrix{1&0&0&0\cr 0&1&0&0\cr 0&0&0&1\cr 0&0&1&0\cr}
          \right)\\
 &=& \ket{0}\bra{0} \one + \ket{1}\bra{1} \sigma_x
\end{eqnarray}
This can be rewritten in terms of exponential operators (see general
discussions by Somaroo et al. for more details
\cite{somaroo:qc1998a,price:qc1999a,price:qc1999b})
\begin{equation}
U_{cnot}=e^{-i{\pi\over 4}}e^{i\slb{\sigma_x}{b}{\pi\over 4}}e^{i\slb{\sigma_z}{a}{\pi\over 4}}
e^{-i\slb{\sigma_z}{a}\slb{\sigma_x}{b}{\pi\over 4}},
\end{equation}
all of which conveniently commute so that the order of implementation
does not matter.  The first term is an overall phase which we may
neglect.  The second term is a $90^\circ$ RF
pulse on the b-spin, and the third a phase shift of the a-spin.  We
have already seen how these can be applied.  The last term is the one
of interest, a two body interaction which induces a conditional
rotation of the b-spin.  The only two body Hamiltonian available in
liquid state NMR is the scalar coupling $\slb{\sigma_z}{a}\slb{\sigma_z}{b}$.
We may once again use AHT to
transform this Hamiltonian into a propagator of the desired form, in
this case by sandwiching a period of spin evolution under the scalar
coupling with a pair of $90^\circ$ rotations of the b-spin about the y
axis,
\begin{equation}
e^{-i\slb{\sigma_z}{a}\sigma_x^(b){\pi\over 4}}=
e^{-i\slb{\sigma_y}{b}{\pi\over 4}}
e^{-i\slb{\sigma_z}{a}\slb{\sigma_z}{b}{\pi\over 4}}e^{i\slb{\sigma_y}{b}{\pi\over 4}}
\end{equation}
Note that the internal Hamiltonian is always present, and while it is
often neglected during non-selective pulses it must be included during
selective pulses.  In addition, the chemical shifts of other spins 
and couplings other
than between spins $a$ and $b$ should be refocused during the
coupling evolution period.  This is normally accomplished via a series of
$180^\circ$ rotations similar to those used during the no-operation gate.
A pulse sequence which accomplishes this for three spins is shown
in Fig.~\ref{figure:cnot}.

\subsection{Measurement}
\label{subsection:measurement}

NMR measurements are weak (as opposed to strong projective
measurements), therefore we can follow the evolution of the spin
magnetization over time with only small distortions being introduced
by the detection coil~\cite{bloembergen:qc1954a}. 
Because the coil is classical we may treat the coupling
between it and the spin system as an example of magnetic induction.
The measured electro-motive force (emf) is then,
\begin{equation}
emf= -{d\over dt}\int {B_1 \cdot M(t) \over |B_1| } d\Omega,
\end{equation}
where $B_1 \over |B_1|$ is the normalized intensity of the RF field generated by
the detection coil, $M(t)$ is the macroscopic local spin magnetization,
and the integral is over the entire volume.  Since the RF coil generates a
dipolar field, only those portions of density matrix that are dipolar
and oriented along the field $B_1$ will couple.
The signal is
measured over time by currents induced by the rotating bulk magnetization
in a transverse coil tuned to the resonance frequency, so that the measured signal
can be given as a trace
\begin{equation}
\textrm{obs}(t)= Tr\Big[\sum_j \slb{\sigma_+}{j} \rho \Big],
\end{equation}
where $\slb{\sigma_+}{j} =\slb{\sigma_x}{j} +i\slb{\sigma_y}{j}$.
The density matrix $\rho$ is modulated in time by the internal
Hamiltonian, and is implicitly shifted to the rotating frame associated with the resonance frequency.
As a result, in product operator terms, the following six operators
\begin{eqnarray}
\sigma_x^a  \one \one&,&\sigma_y^a  \one \one
\\ \one 
\sigma_x^b \one&,&\one \slb{\sigma_y}{b} \one \\
\one \one \sigma_x^c&,&\one \one \sigma_y^c \\
\end{eqnarray}
are directly observable for the three spin system.

Other product operators are indirectly observable since the NMR
signal is detected over time, and with normally evolve under the influence of the
internal Hamiltonian. Thus the directly observed product operators
for spin $a$ shows a modulation from coupled spins via the action of the scalar coupling.
For example,
\begin{eqnarray}
\slb{\sigma_x}{a}  \slb{\sigma_z}{b} \one &\mathop{\longrightarrow}
\limits^{ {\pi\over 2} J_{ab}\slb{\sigma_z}{a}\slb{\sigma_z}{b}}& 
\cos[{\pi\over 2} J_{ab} t] \slb{\sigma_x}{a} 
\slb{\sigma_z}{b} \one +
\sin[{\pi\over 2}  J_{ab} t] \slb{\sigma_y}{a}  \one \one.
\end{eqnarray}
As a result, an additional eighteen terms which evolve in and out of
directly observable terms are indirectly observable. This includes
$\slb{\sigma_x}{a}\slb{\sigma_z}{b}$, but not, for example,
$\slb{\sigma_x}{a}\slb{\sigma_x}{b}$ which does not evolve under the scalar coupling Hamiltonian.

The decoupling methods from
Sec.~\ref{subsection:decoupling} suppress the scalar coupling to the
decoupled spin and thus, so-called anti-phase states such as
$\slb{\sigma_x}{a} \slb{\sigma_z}{c}$ will not be observed in
the measurement.  This is consistent with considering decoupling
as a tracing operation.

For deterministic algorithms, it is sufficient to measure a single
spin.  For small Hilbert spaces, however, it is often convenient to
measure the full density matrix when exploring the performances of a
QP.  For most QPs where the measurement involves collapsing into a
measurement basis, reading out the density matrix involves a
tomographic procedure of repeated measurements as a function of a
rotation of the system into the measurement basis.  The density matrix
is then reconstructed by inverting the measurement results.
In NMR, since the measurement is weak, all transverse dipolar degrees of
freedom can be read out directly and in one step.  The result is that
state tomography can be achieved in fewer steps and the density matrix
reconstructed via spectral fitting.  The method involves appending
readout pulses to the end of the calculation to rotate non-observable
operators into observable ones.  For the three qubit
system under discussion, seven measurements (with different readout
pulses) are sufficient to completely characterize the density matrix.
Following the measurement of a complete set of spectra, these may be
fitted to extract the complete description of the density matrix.

\subsection{Non-unitary operations}

A desirable operation for a QP is a strong projective measurement.  In
cases where the projective measurement does not contribute toward the
final answer, such as is the case in error correction procedures, only
its effect on the state, not its outcome, is required.  This effect
can be mimicked by decohering the off-diagonal elements in a
measurement basis.  This can be accomplished through natural decoherence after having rotated the spins in the desired basis~\cite{nielsen:qc1998a}, or by the action of a magnetic field gradient followed by
molecular diffusion to remove all recoverable spatial correlations.
Thus ensemble dynamics makes it is possible to mimic 
strong measurement processes, although unfortunately this cannot be used as
a means of repolarizing the system since a mixed state results.

\subsection{Pseudo-pure states}
\label{subsection:pps}

One example where controlled decoherence is necessary is the
preparation of pseudo-pure states.  At room temperature 
the NMR spin system is in a highly mixed state since $\Delta E/kT$ 
is of order $10^{-5}$.  The sample consists
of about $10^{18}$ molecules and the state must be described by a density matrix which at
equilibrium for a single nuclear species is
\begin{equation}
\rho_{eq}={1\over Z} e^{-H/kT}\approx {1\over Z} \one +\epsilon_1\sum_i \sigma^i_z
\end{equation}
where $Z$ is a normalization factor and recall that $\omega/J\approx 10^6$ 
so to a good approximation the bilinear term in the
Hamiltonian does not influence the populations.  As long as relaxation 
is unimportant the
identity portion of the density matrix is a constant, and is
unobservable.  In a homogeneous magnetic field, we have no useful control over the identity part of the
equilibrium density matrix, so NMR methods rely on pseudo-pure states
whose dynamics are equivalent to those of pure states with respect to
weak measurements. An example of a pseudo-pure state is given by
\begin{equation}
\rho_{pp} =  {1\over Z} \one +\epsilon_2\prod_i \kets{0}{i}\bras{0}{i},
\label{eq:idealpp}
\end{equation}
which has the property that its traceless part is up to a scalar
constant equal to that of a pure state. The scalar constant is the
intensity of the pseudo-pure state.  Methods for creating pseudo-pure states on all spins have intensity decreasing exponentially
with the number of qubits.  This is why for general quantum computing,
liquid state NMR is limited to about ten spins.

\begin{figure}
\begin{center}
\mbox{
\includegraphics[height=1.8in,width=4.2in]{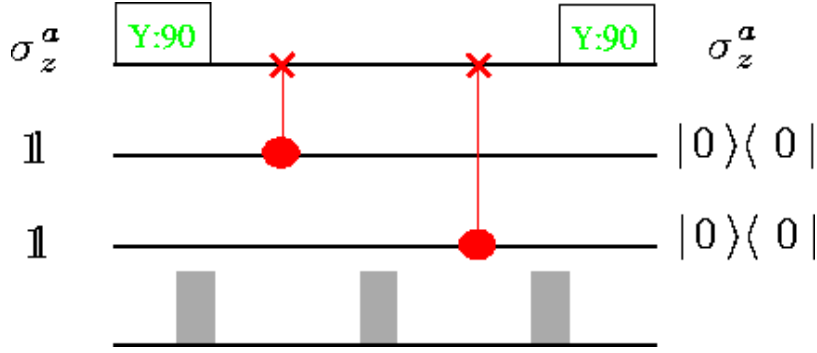}
}
\end{center}
\caption{Pulse sequence to create a pseudo-pure state from the state
$\slb{\sigma_z}{a}\one\one$.
The grey filled rectangle in the bottom line correspond to gradient pulses.
The relative area of the three gradient pulses are shown in the figure.}
\label{figure:ppstate}
\end{figure}

Instead of generating the pseudo-pure state of Eq.~\ref{eq:idealpp},
it is usually simpler to create
\begin{equation}
\rho_{xpp} = \lambda \one + \epsilon_3\slb{\sigma_x}{0}\prod_{i\geq 1}\kets{0}{i}\bras{0}{i}
\end{equation}
This state can either be used as a standard pseudo-pure state on one
less spin (the measurement needs to be somewhat modified), or in
many cases it can serve directly as one of the inputs in benchmarking
an algorithm involving coherent processing of one
qubit~\cite{knill:qc1999a}. Decoherence operations such as those
available with gradients yield particularly simple means for creating
states of this form. A procedure which accomplishes this for three spins
is given in Fig.~\ref{figure:ppstate}.

We can again use AHT to follow the gradient Hamiltonian in the toggling frame
of the C-not pulses for the sequence of Fig.~\ref{figure:ppstate}.
There are three intervals, and the evolution is by a
position dependent rotation about the z axis.
\begin{center}\begin{tabular}{c|ccc|c|c}
& \ 1 \  & \ 2\ &\ 3 &\ average
\ &\ $2\theta_1=\theta_2=2\theta_3=\theta$ \cr
\hline
$E_+^aE_+^bE_+^c$ & \ $\theta_1$ \  & \ -$\theta_2$\
  &\ $\theta_3$&\ $\theta_1-\theta_2+\theta_3$&\ 0\cr
$E_+^aE_+^bE_-^c$ & \ $\theta_1$ \  & \ -$\theta_2$\
  &\ $-\theta_3$&\ $\theta_1-\theta_2-\theta_3$&\
$-\theta$\cr
$E_+^aE_-^bE_+^c$ & \ $\theta_1$ \  & \ $\theta_2$\
  &\ $-\theta_3$&\ $\theta_1+\theta_2-\theta_3$&\
$\theta$\cr
$E_-^aE_+^bE_+^c$ & \ $\theta_1$ \  & \ $-\theta_2$\
  &\ $\theta_3$&\ $\theta_1-\theta_2+\theta_3$&\
0\cr
$E_+^aE_-^bE_-^c$ & \ $\theta_1$ \  & \ $\theta_2$\
  &\ $\theta_3$&\ $\theta_1+\theta_2+\theta_3$&\
$2\theta$\cr
$E_-^aE_+^bE_-^c$ & \ $\theta_1$ \  & \ $-\theta_2$\
  &\ $-\theta_3$&\ $\theta_1-\theta_2-\theta_3$&\
$-\theta$\cr
$E_-^aE_-^bE_+^c$ & \ $\theta_1$ \  & \ $\theta_2$\
  &\ $-\theta_3$&\ $\theta_1+\theta_2-\theta_3$&\
$\theta$\cr
$E_-^aE_-^bE_-^c$ & \ $\theta_1$ \  & \ $\theta_2$\
  &\ $\theta_3$&\ $\theta_1+\theta_2+\theta_3$&\
$2\theta$\cr
\end{tabular}\end{center}
where $E_\pm^i =(\one \pm\slb{\sigma_z}{i})/2$.
Putting this all together and writing the gradient Hamiltonian out
explicitly, the averaged gradient Hamiltonian introduced by the above
sequence is,
\begin{equation}
\bar H_{grad}=\gamma {\partial B_z\over \partial z} z\slb{\sigma_z}{a}
( -\slb{E_+}{b}\slb{E_-}{c} + \slb{E_-}{b}\slb{E_+}{c} + 2\slb{E_-}{b}
\slb{E_-}{c} )
\end{equation}
The effect of the dependence on $z$ and subsequent diffusion
is that the input state projects to the desired pseudo-pure state.

There are many other approaches to create pseudo-pure states,
including temporal averaging schemes, and logically labeling the state
dependent on another spin
\cite{stoll:qc1977a,suter:qc1986a,knill:qc1997b,chuang:qc1997a,knill:qc1999a}.
The polarization of all of these falls off exponentially with the number
of qubits. 

\subsection{Achievements of liquid state NMR QIP}

One reason why NMR QIP has had such a head start over other approaches
to QIP is that NMR spectroscopy has been used for coherent
manipulations for decades. Indeed much work on NMR is directed towards
coherently controlling quantum states so as to extract specific
chemical information from complex systems.  This has led the NMR
community to develop powerful analytic tools for understanding spin
dynamics. 

In the weak coupling limit, where qubits are most easily implemented
in terms of nuclear spins, the product operator
formalism~\cite{sorensen:qc1983a,vanderven:qc1983a,packer:qc1983a}
provides a complete description of the dynamics.  Not only is this
extremely useful experimentally, but it also exactly describes the
desired dynamics of a quantum computer.  Of course for NMR spectroscopy the
focus is on single-spin operations and scalar couplings, and for QIP
it is on single qubit operations and C-not gates.  In the case of more
complex spin systems, either with non-negligible strong coupling or
with spins $>{1\over 2}$, the dynamics is often expanded in a
fictitious spin $1\over 2$ formalism~\cite{abragam:qc1961a}, which
essentially is a mapping of the physical system into qubits in the
manner suggested for quantum simulation outlined in
Fig.~\ref{figure:simphys}.  Beyond a certain complexity it is far too
difficult to map the dynamics into distinguishable qubits and the
focus turns to thermodynamic properties of the spin system via the
concepts of spin temperature~\cite{goldman:qc1970a}.  A major 
challenge of QIP is to use these larger systems for information
processing. 

We have seen repeated examples of how average Hamiltonian theory is
related to quantum simulation and extremely useful for choosing
specific implementations of quantum gates.  Although limited space has
not permitted a detailed discussion, similar control may be gained over
non-unitary superoperators through the use of average Liouvillian
theory \cite{ghose:qc1999a,tseng:qc2000a}.  In general relaxation is
understood through superoperator theories~\cite{jeener:qc1982a}, and
these ideas are just now being combined with QIP in the context of
noiseless subspaces~\cite{zanardi:qc1997a,lidar:qc1998a} and
subsystems \cite{knill:qc1999b}.  Interestingly the concepts
of quantum error correction~\cite{shor:qc1995b} do not seem to have
been anticipated in even a primitive sense by the NMR community.  This
is perhaps an example of where complex ideas can only be reached
through a formal information theory approach.

The same features that make QIP possible, namely the ability to strongly modulate
the spin dynamics on a time scale short compared to decoherence, has
lead to a wealth of NMR methods aimed at specific chemical and
physical problems.
Many of the experimental techniques that have made the NMR
implementations of QIP proceed so quickly are well established in NMR.
There are very robust methods for creating selective
pulses~\cite{warren:qc1980a,redfield:qc1975a,forsen:qc1963a,morris:qc1978a},
and these have reached a very high precision for the purposes of
magnetic resonance imaging.  Almost every NMR experiment today makes
use of some form of coherent control: (1) to transfer polarization for
sensitivity enhancement~\cite{morris:qc1979a,doddrell:qc1982a}, (2) to
transfer coherence for correlation
experiments~\cite{bax:qc1981a,aue:qc1976a}, (3) to selectively
suppress coherence for solvent signal elimination and to limit the
dynamic range of the signal~\cite{bodenhausen:qc1984a}, and (4) to
select multiple quantum pathways to observe specific desired
subsystems of spins~\cite{burum:qc1980a,warren:qc1979a}.  Such control
is also used to tailor the dynamics: (5) for decoupling undesired
spins~\cite{levitt:qc1981a,waugh:qc1982a}, (6) for recoupling spins whose
chemical shift differences are greater than the coupling between
them~\cite{chingas:qc1981a} and (7) to selectively remove
dipolar~\cite{waugh:qc1968a} or even
quadrupolar~\cite{mueller:qc1990a} interactions.  Most of these
methods have found applications in QIP, and the precision of control
required by QIP implies that we will have to make use of all possible
means of coherent control.

NMR spectroscopy has not exclusively focused on chemical and medical applications,
it has also been used as a probe of quantum dynamics.
Many experiments have been carried out to explore the structure of
quantum dynamics.  The field has struggled with the question of bounds
on the efficiency of transformations such as polarization 
transfer~\cite{sorensen:qc1989a} and on finding means to achieve these
experimentally \cite{sorensen:qc1990a}.  QIP-like explorations
include interferometric measurement of spinor
dynamics~\cite{stoll:qc1977a} (the creation of logically labeled
pseudo-pure states were needed for these) and explorations of
geometric phase~\cite{suter:qc1986a}.  There is also a recent
NMR implementation of geometric phase as a controlled
gate~\cite{jones:qc2000a}.

There has also been a series of explorations of quantum complexity
for systems of dipolarly coupled spins conducted in both liquid
crystals~\cite{saupe:qc1965a} and the solid
state~\cite{munowitz:qc1987a}.  It is the realization that NMR studies
of dipolarly coupled spins provides a test bed for complex quantum
dynamics that has led Chuang~\cite{yannoni:qc1999a} and us to propose
next generation QIP devices built from liquid crystals and solids,
respectively.  The solid state also offers an ideal test bed for
larger studies of multi-body dynamics and of the transition from
quantum to classical behavior~\cite{zhang:qc1998a}.

In this contribution we  cannot do justice to all the NMR QIP
experiments that have been realized and certainly not to 50 years of
NMR spectroscopy.  Clearly without the accumulated knowledge of NMR
spectroscopy theory, methods and instrumentation, the field of NMR QIP
would not be so advanced.  Fortunately there are excellent and detailed
descriptions of NMR~\cite{abragam:qc1961a,slichter:qc1963a,haeberlen:qc1976a,ernst:qc1994a,pines:qc1994a}
The following table summarizes the NMR QIP experiments published to date.

\begin{center}\begin{tabular}{|c|l|l|l|}
\hline
\# of qubits     &  Algorithms 
   &  Year & Reference  \cr\hline\hline
2     &  Gates  
      & 1996  & 
\cite{cory:qc1996a,chuang:qc1997b,cory:qc1998b,collins:qc1999a,price:qc1999a,marjanska:qc2000a,jones:qc2000a}
 \cr\hline 
&  Database Search  
  & 1998 & \cite{jones:qc1998b,chuang:qc1998c,yannoni:qc1999a} \cr\hline
  &  Deutsch-Josza 
 & 1998 &  \cite{jones:qc1998a,chuang:qc1998a} \cr\hline
&  Quantum Simulation  
  & 1999 &  \cite{somaroo:qc1999a}  \cr\hline
& Quantum Detecting Code  
 & 1999 &  \cite{leung:qc1999b} \cr\hline
& Quantum Fourier Transform  
 &1998 & \cite{weinstein:qc2000a,fu:qc1999a}\cr\hline
& Dense Coding  
 &1998 & \cite{fang:qc1999a} \cr\hline
& Quantum Detecting Code
 &1999 & \cite{leung:qc1999b} \cr\hline\hline
3    & GHZ state  
  & 1997 &  \cite{laflamme:qc1997a,nelson:qc2000a}\cr\hline
& Quantum Error Correction  
 & 1997 &  \cite{cory:qc1998a} \cr\hline
& Quantum Teleportation  
 & 1997 &  \cite{nielsen:qc1998a}\cr\hline
& Deutsch-Josza  
 & 1998 &  \cite{dorai:qc1999a,linden:qc1998a} \cr\hline
& Quantum Simulation 
  & 1999 &  \cite{tseng:qc2000a} \cr\hline
& Quantum Fourier Transform  
 &1998 & \cite{weinstein:qc2000a}\cr\hline
& Quantum Eraser  
 &1998 & \cite{tek:qc2000a}\cr\hline\hline
4 & C$^3$-not Gate  & 1999 & \cite{price:qc1999a} \cr\hline
5 & Deutsch-Josza  & 1999 & \cite{marx:qc1999a} \cr\hline
6    &  Decoupling 
  & 1998 &  \cite{linden:qc1999a} \cr\hline
7    & Benchmark 
  & 1999  &  \cite{knill:qc1999a}  \cr\hline
\end{tabular}\end{center}

\section{Limitations of liquid state NMR QIP}
\label{section:nmr-limits}

Although liquid state NMR is extremely useful to demonstrate
coherent control and bring theoretical quantum computation
to the laboratory it has the following limitations:  
\begin{itemize}
\item[1)] The non-scalability of pseudo-pure state preparation.  As the number of
qubits increases, the signal resulting from the pseudo-pure state
preparation decreases exponentially. This decay quickly becomes
prohibitive and renders quantum error correction ineffective
since it requires ancillas in pseudo-pure states.
\item[2)] The large ratio of the gate time over the decoherence time (of
order $.01$).  This is sufficient for proof-of-principle experiments but
too small for computationally interesting algorithms.
\item[3)] The difficulty of resetting qubits.  In order to iterate
quantum error correction, ``fresh'' qubits must be supplied.  This can
be achieved by either  having initially a large supply that
remains perfectly polarized or by refreshing the qubits by setting
them back to a fiducial state. Liquid state NMR does not seem to
provide a mechanism for resetting qubits at will.
\item[4)] As we enlarge the number of qubits the spread of Larmor
frequencies (limited by the chemistry) will have to increase
accordingly in order to preserve the same degree of control. Although this
can be circumvented, doing so is difficult~\cite{lloyd:qc1993a}.
Unfortunately nature limits the range of chemical shifts to about 200
ppm of the resonance frequency for carbon, sufficient to distinguish
only about 20 carbon nuclei for QIP purposes at 
reasonable field strengths and operation times.
\end{itemize}

The first generation of QPs based on liquid state NMR has been very
useful as a first step in a program towards a quantum computer.  But faced
with the above limitations we now propose to move to a solid state NMR-based device which we believe will carry QIP beyond 10 qubits.

\section{Introduction to solid state NMR}

Before presenting the next generation of NMR quantum information
processors we quickly survey the essential elements distinguishing
liquid and solid state NMR. The solid state offers four main
advantages for QIP.  First, the system can be highly polarized
increasing the sensitivity necessary to read out the results of
computations involving many qubits.  Second, the decoherence rates can be made
slower. Third, couplings between spins are higher, thereby permitting
both faster and more accurate operations, so that algorithms
of much greater complexity can be implemented.
Finally, there are mechanisms which can be used
for dynamically resetting qubits, thus making it possible
to remove information from a system and to implement error-correction
efficiently.

From a spectroscopy point of view the main differences between
liquid and solid state NMR are:
\begin{itemize}
\item[1)]  The dipolar couplings are resolved.
\item[2)]  Spin diffusion contributes significantly to the dynamics.
\item[3)]  The chemical shift needs to be described by a tensor.
\item[4)]  The spin lattice relaxation times are typically very long.
\item[5)]  The transverse (phase) decoherence time is dominated by spin-spin
interactions which can be refocused by special purpose RF pulse sequences.
\item[6)]  Methods are known for creating very high non-thermal polarizations.
\end{itemize}

In the liquid state the interaction between distinguishable nuclear
spins is mediated through electrons as an indirect interaction or
J-coupling.  In the solid state, the direct dipole-dipole coupling is
time independent and thus observed.  The internal Hamiltonian in the
solid state has the form,
\begin{eqnarray}
H_{solid} = \sum_i {\omega_i(\theta,\phi)\over 2}\slb{\sigma_z}{i} + 
\sum_j\sum_k {\gamma_j\gamma_k \over 4r^3_{j,k}} (3 \cos^2\theta_{j,k} -1)
\sigma^j_z \sigma^k_z \nonumber \\
-\sum_j\sum_k {\gamma_j\gamma_k \over16 r^3_{j,k}} (3 \cos^2\theta_{j,k} -1)
(\sigma_+^j\sigma_-^k + \sigma_-^j\sigma_+^k)
\end{eqnarray}
where $\theta$ and $\phi$ are the Euler angles associated with the
orientation of the molecule in the external magnetic field and
$\theta_{j,k}$ and $r_{j,k}$ are the angle and magnitude of the
difference vector between nuclei $j$ and $k$ with respect to the
external field. 

In the homonuclear case the ``flip-flop'' $(\sigma_+\sigma_- + \sigma_-\sigma_+)$ term
is important and results in spin diffusion, while in the
heteronuclear case this term is non-secular and may be dropped from
the Hamiltonian.  The heteronuclear case is thus very similar to the
liquid state situation described above, except that both the dipolar
couplings and chemical shifts have an orientation dependence
(relative to the external magnetic field).  It will therefore be
important to use single crystals so that there is a unique mapping of
the spin system to the resonance frequency, or, alternatively to apply
magic angle sample spinning \cite{andrew:qc1959a} to remove the
orientation dependence.  If a single crystal is used, then the sample
must crystallize with only one molecule per unit cell.

In the homonuclear case the dipole-dipole coupling leads to a mixing
of states that often prevents the chemical shift term from being
useful for spectrally separating chemically distinct spins.  In
addition, the flip-flop term leads to an apparent short transverse
decoherence rate.  The flip-flop dynamics is however perfectly coherent and
should be viewed in the light of recording the history of spin
evolution.  However, it has the reputation of being decoherent because
it is normally studied in samples where the homonuclear chemical
shift differences are not resolved and thus the observed Hilbert space is that
resulting from a trace over what are in fact distinct spins.  Via this
trace operation the coherent flip-flop dynamics are artificially
mapped into an apparently non-unitary evolution. For most samples the
true transverse decoherence rates are extremely challenging to measure
requiring suppression of the dipolar interaction over the duration
of the measurement~\cite{cory:qc1990a}.

\begin{figure}
\begin{center}
\mbox{
\includegraphics[height=1.5in,width=4.8in]{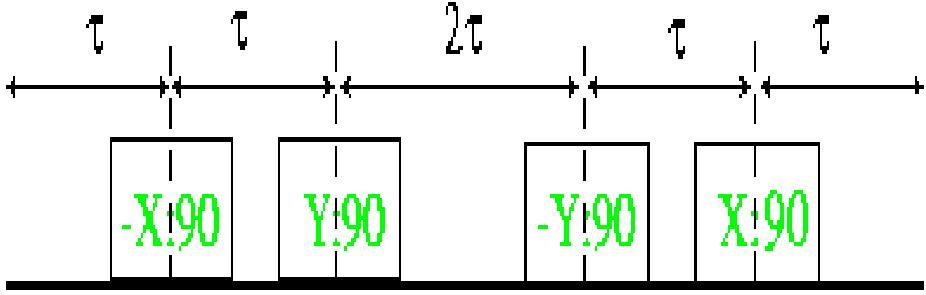}
}
\end{center}
\caption{WAHUHA pulse sequence to reduce the dipolar interaction
between spins. The sequence is applied to all nuclei.}
\label{figure:wahuha}
\end{figure}

To control and observe the spin dynamics in the presence of dipolar
coupling one needs to employ multiple-pulse coherent averaging which
independently rescales the dipolar and chemical shift terms of the
Hamiltonian in the AHT.  A simple and descriptive example is the
WAHUHA sequence~\cite{waugh:qc1968a}, the effects on the terms of the Hamiltonian in the
toggling frame are,
\begin{center}\begin{tabular}{l|@{\protect\rule[8pt]{0pt}{15pt}\ }ccccc|l}
term &  1   &  2 & 3 & 4 & 5 
& average \cr
\hline
$\slb{\sigma_z}{a}$ &  $\slb{\sigma_z}{a}$   &  $\slb{\sigma_y}{a}$
  & $\slb{\sigma_x}{a}$& $\slb{\sigma_y}{a}$& $\slb{\sigma_z}{a}$
&${1\over 3} (\slb{\sigma_x}{a}+\slb{\sigma_y}{a}+\slb{\sigma_z}{a})$\cr
$\slb{\sigma_z}{b}$ &  $\slb{\sigma_z}{b}$   &  $\slb{\sigma_y}{b}$
  & $\slb{\sigma_x}{b}$& $\slb{\sigma_y}{b}$& $\slb{\sigma_z}{b}$
&${1\over 3} (\slb{\sigma_x}{b}+\slb{\sigma_y}{b}+\slb{\sigma_z}{b})$\cr
$\slb{\sigma_z}{a}\slb{\sigma_z}{b}$ &  $\slb{\sigma_z}{a}\slb{\sigma_z}{b}$   
& $\slb{\sigma_y}{a}\slb{\sigma_y}{b}$& $\slb{\sigma_x}{a}\slb{\sigma_x}{b}$
& $\slb{\sigma_y}{a}\slb{\sigma_y}{b}$&$\slb{\sigma_z}{a}\slb{\sigma_z}{b}$ 
&${1\over 3}\slb{\sigma}{a}\cdot\slb{\sigma}{b}$\cr
$\sigma^a\cdot\sigma^b$ &  $\sigma^a\cdot\sigma^b$   
& $\sigma^a\cdot\sigma^b$& $\sigma^a\cdot\sigma^b$
& $\sigma^a\cdot\sigma^b$&$\sigma^a\cdot\sigma^b$ 
&$
\sigma^a\cdot\sigma^b$\cr
\end{tabular}\end{center}
From the above scheme it can be seen that the homonuclear dipolar coupling averages to
\begin{equation}
\propto \sigma^a\cdot\sigma^b - 
3 \langle \slb{\sigma_z}{a}\slb{\sigma_z}{b}\rangle =0,
\end{equation}
though the chemical shift term remains.  The averaged chemical shift Hamiltonian is now oriented along the $\vec x + \vec y + \vec z$ direction. In practice more complex sequences are used that include finite pulse width effects and also higher order terms in the Magnus expansion.

Related sequences such as the magic echo~\cite{rhima:qc1970a} rely on
the convenient property that in a strong spin locking field  the
dipolar Hamiltonian is scaled by a factor $-0.5$.  Thus under spin
locking conditions the dipolar Hamiltonian can be refocused, because the
coherent part of the dynamics appears to run backwards.

The spin-lattice relaxation times of diamagnetic crystalline solids
below their Debye temperature are generally very long (of the order of
hours) and are frequently dominated by the presence of
defects \cite{bloembergen:qc1949a}. A common defect is a color center
containing free electrons. Since the electron spin correlation time is
short, this acts as a locally fluctuating field that decoheres nearby
nuclear spins.  This local spin magnetization gradient is then carried
through the lattice by the flip-flop term of the homonuclear dipolar coupling.  Indeed it has been
observed that in the presence of off-resonance excitation (to quench
spin diffusion) the $T_1$ becomes longer, and this can be attributed to
the fact that relaxation is mediated by magnetization transport and
without spin diffusion there is no way for the nuclear magnetization
to reach the electron relaxation sinks.

Stable free electrons like those in color centers are also valuable
for providing a source of polarization.  The electron magnetic moment
is about 1800 times that of the proton, so at $4^\circ K$ the electron
spins are highly polarized.  This polarization can be transferred
coherently or via anisotropy in decoherence to the local nuclear
spins, from which spin diffusion can carry the polarization to the
more distant portions of the sample.  The electron relaxation times
are short, and so the effective heat capacity of the electron spin
bath is high enough to polarize many nuclear spins.

\section{Next generation QIP based on solid state NMR}
\label{section:nextqip}

\subsection{General scheme}

\begin{figure}
\begin{center}
\includegraphics{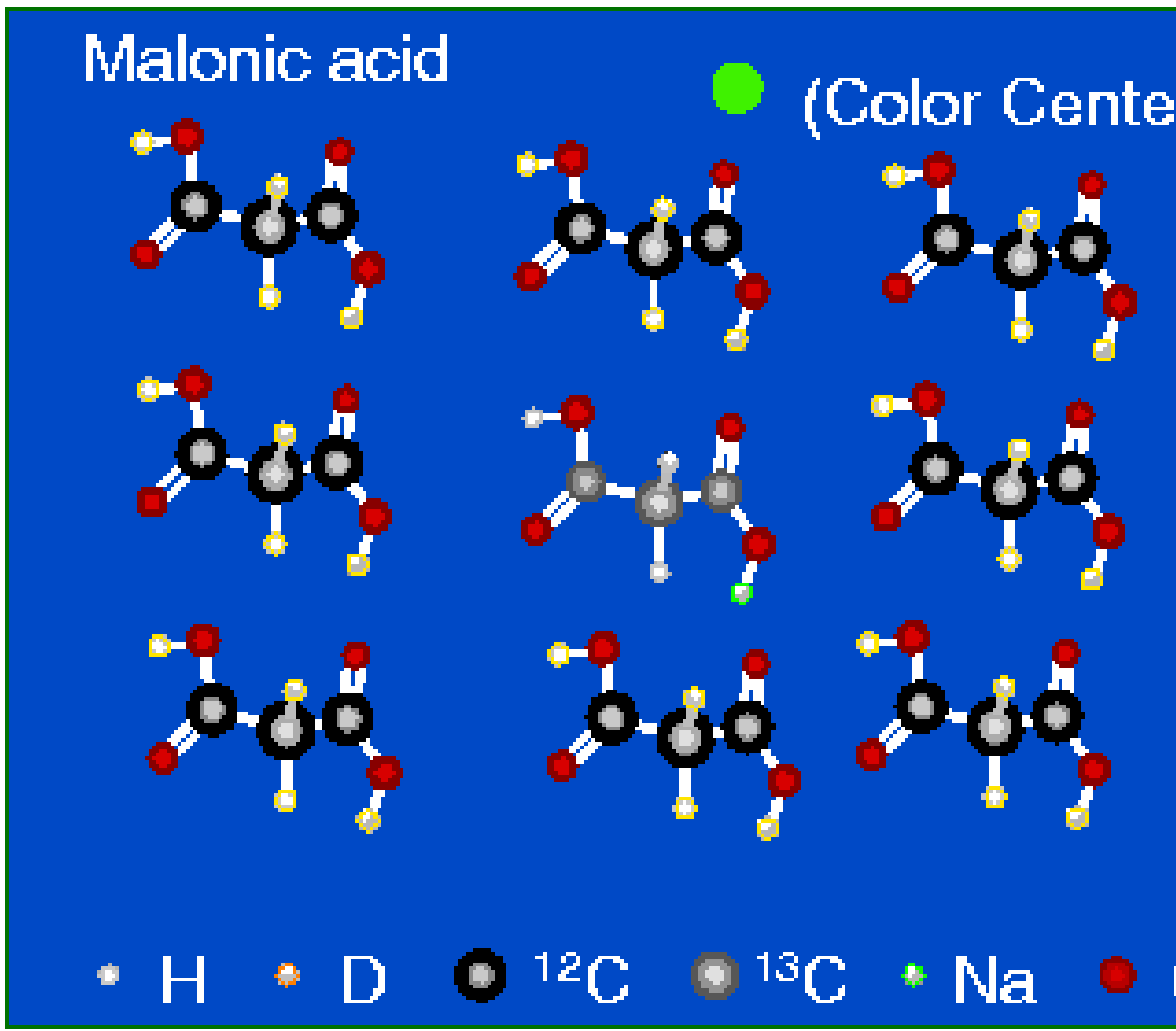}
\end{center}
\caption{Solid state NMR quantum processor using malonic acid for
illustration. An idealized lattice of the molecules is shown.  One of
the molecules has $^1H$, $^{13}C$ and one sodium atom, whose nucleus
serves as a resettable qubit. This molecule is immersed in a
background lattice of deuterated and $^{13}C$ depleted crystal of the
same acid with no sodium. Color centers are included to generate large
non-thermal polarizations.}
\label{figure:malonic}
\end{figure}

To build up to a prototype QP capable of non-trivial quantum
algorithms, simulations, and decoherence studies, we propose a next
generation of NMR QPs based on solid state.  In this section we first
introduce the general scheme, then give a detailed description of the
key steps and components.  A large scale QP can be envisioned on the
basis of the following device consisting of an ensemble of
microscopic, identical QPs.  The heart of the system is an n-qubit QP
composed of a set of spin $1\over 2$ nuclei in a precisely
engineered arrangement associated with the chemistry of a designed
molecule. Individual QPs are held and aligned in a lattice in
which the only active nuclear spins are deuterium nuclei.
To achieve uniformity of the lattice, it is convenient to use an
isotopic variant of the molecule that is used for the QP.  If the
spins used for QIP are given by $^1H$ and $^{13}C$, a useful variant
is fully deuterated and unlabeled.  The entire ensemble is thus a
single crystal of the deuterated molecules with a very dilute
concentration of suitably labeled versions. The concentration of QPs is
kept low enough that the individual QPs do not communicate with one
another.  Spectroscopic studies of such molecular preparations are
well known, and are useful when the goal is to obtain well resolved
proton spectra from complex molecules \cite{snyder:qc1973a}. Since
the individual QPs are in close proximity to the deuterated bath, an
important component of this approach is to completely decouple the
deuterons from the spin $1\over 2$ nuclei via multiple-pulse sequences.
Methods for such decoupling sequences are well known~\cite{pines:qc1976a}.
An illustration of our proposal is in Fig.~\ref{figure:malonic}.

The molecules that make up the individual QPs are designed so that the
dipolar couplings between spins remains strong (of the order of kHz)
while the spectral separation between nuclei is sufficiently large for
the qubits to remain individually addressable.  This is facilitated by the fact
that in the solid state both the chemical shift and dipolar coupling 
retain their anisotropy and hence reorientation of the sample relative 
to the external magnetic field provides a convenient way for
controlling their relative strengths.  An important goal is to ensure
that the spectral separation between any two qubits is appreciably
larger than their dipolar coupling.  This should be achievable in
principle, due to the fact that the dipolar coupling goes to zero when
the inter-nuclear vector is at the magic angle to the applied magnetic
field.

An additional simplification occurs at high polarizations in that the
flip-flop term of the Hamiltonian is quenched and thus configurational
broadening of the resonance line is avoided both from within the QC
and with the deuterated lattice (external to the QC) which, as stated,
is actively decoupled.  

The deuterated lattice can include color centers (for
example, stable free electrons can be introduced through gamma
irradiation). They are used to generate large non-thermal
polarizations in the nuclear spins through the well established method
of dynamic nuclear polarization (DNP)~\cite{wind:qc1985a}.  The spin
$1\over 2$ nuclei that make up the QP are polarized in a two step
process. First, the deuterium lattice is polarized by transfer of the
free electron polarization. Then this polarization is transferred to
the QP's spins via cross polarization~\cite{hartman:qc1962a}.  Since
the heat capacity of the lattice is much greater than that of the
dilute molecules of the QP this transfer will polarize the QP close to
the initial polarization of the electrons.

To increase the DNP signal enhancement the system is kept at $-8^\circ K$, at which temperatures the
electron spins are nearly in a pure state.  The system is also arranged so that
the QPs are far enough apart that they do not directly see the color
centers, which under normal circumstances would be the primary
source of decoherence (the crystal is well below its Debye
temperature).  At such low temperatures the spin-spin interactions
are more important than the phonon motions and the only fluctuating
magnetic fields arise from the lifetimes of the z-component of
the electron spin, which has a relatively short spin-lattice
relaxation time.  The double quantum decoupling method mentioned
earlier also decouples the deuterium
spins from one another so that spin transport through the lattice is
quenched and the color centers no longer play a significant relaxation
role~\cite{mavlonazarov:qc1992a}.  Thus the decoherence times of the
QPs will be very long (of order $10^3$ seconds).  Notice that the electrons
play no role in the computation and their short decoherence time is in
fact a benefit for rapid state preparation without in any way limiting
the achievable computational complexity.

In future implementations it may be feasible to include an erasable
quantum register.  This permits the complete purification of states
through polarization pumping~\cite{schulman:qc1998a}, the efficient
implementation of quantum error-correction codes, and provides a
source of decoherence to mimic the effect of strong measurements.
This register could be composed from a set of quadrupolar spins in the
molecule of a QP that are engineered to be coupled to the QP's qubits.  Quadrupolar nuclear spins couple to the local
electric field gradient and thus are sensitive to acoustic
modulations~\cite{proctor:qc1955a} which enhance decoherence rates.  An acoustic pulse can thus decohere the erasable register,
while leaving the spin $1\over 2$ nuclei in the QP in any given state.
The erasable register must then be repolarized.
This use of erasable ancilla results in even small QCs being quite
powerful since quantum resources need not to be borrowed for
non-computational tasks.

\subsection{Requirements}

The five requirements of a physical quantum computer are explored
below.  We hope to convince the reader that the requisites for
building a 
functional prototype quantum computer are within reach through an
amalgamation of solid state NMR methods.  At the same time we caution
that the proposed QIP device will function on 10 through perhaps 30
qubits, which while making it by far the largest quantum processor to
date (and possibly just beyond what can presently be simulated on a
classical computer) the system will not be scalable.  We
believe however that it is only through the exploration of such
quantum devices that scalable quantum computers might someday be
realized. We consider each requirement in turn.

\subsubsection{Scalable mapping}

Although the correspondence between qubits and spin $1\over 2$
nuclei is clear in this case, its scaling is in fact limited
by the need for having readily crystallizable
molecules. In principle, adequate addressability of the qubits
is available, if necessary by using quantum cellular automata
implementation methods. We believe that it is possible to
design appropriate molecules for up to perhaps 30 qubits.
A more immediate obstacle to implementing this mapping
is the presence of an internal Hamiltonian which does not
preserve the standard basis for qubits, thus requiring active
manipulation even for preserving the state.
We address the issue of state preservation in the context of
the requirement for implementing a sufficient set of gates.

The complexity of the internal Hamiltonian leads to two complications
for viewing the spin system as a system of qubits. First, the
eigenbasis is no longer known and so there is not a convenient and
stable computational basis, and second the resonances are broadened by
the many spin configurations of equal energy (configurational
broadening).  The second part is of little concern since we must fully
control the internal Hamiltonian to have a QP, but the first point
implies that individual qubits cannot be identified or addressed with
RF pulses as easily as is the case in liquid state. The way out of
the this difficulty is to realize that the flip-flop term does not
commute with the difference of $\sigma_z$ operators, and that provided
a spectral difference (such as that introduced by chemical shifts) can
be made stronger than the dipolar couplings, the dynamics from the
flip-flop term are effectively quenched by second averaging.  Such a
quenching is of course guaranteed in the case of heteronuclear
coupling since the differences in Larmor frequencies are some MHz while
the dipolar couplings at most in the tens of kHz.

For the molecular system to be used in this proposal we need only
concern ourselves with the spins that make up the individual QPs,
since the deuterated bath will be decoupled from these, and
each molecular QP will be well isolated from all other copies.  The
QP molecules should thus have the following properties:
\begin{itemize}
\item[1)] Crystallize easily into large single crystals with only one
molecule per unit cell.
\item[2)] Have high Debye temperatures ($> 100 ^\circ K$).
\item[3)] Be dielectric.
\item[4)] Be composed of a few elements in addition to $^1H$ for which there is an isotope with no spin, and another with spin ${1\over
2}$ (such as $^{13}C$).
\item[5)] Have an orientation in which all spins are individually addressable.
\end{itemize}

The complete chemical shift tensors (with principal axis systems) are
only known precisely for a few molecules since there has been little
motivation to measure these.  However, there are enough to ensure that finding a
10 qubit system with the above properties is only a modest challenge.
We will start by exploring organic acids since they have been widely
studied, and crystallize well.  An illustrative example is malonic
acid~\cite{haeberlen:qc1985a}, which is shown in
Fig.~\ref{figure:malonic}.  An excellent area for appreciating the
complexity of homonuclear dipolar coupled spins is zero field
spectroscopy~\cite{zax:qc1985a}.  Zero-field spectroscopy removes the
symmetry breaking introduced by the applied magnetic field and permits
single crystal like spectra to be acquired for powdered samples.
Pines and coworkers have reported both experiments and simulations of
small spin systems that show the power and complexity of these methods, and
with coupling constants spanning a range of 60 kHz for small organic
molecules~\cite{zax:qc1985a}. 

\subsubsection{State preparation}

The state preparation, while perhaps appearing to be complex, is
actually the simplest part of the protocol.  We take advantage of the
large Zeeman splitting of the electron spins so that we have a highly
polarized quantum bath at liquid Helium temperatures.  This saves us
from having to work at the much more challenging mK temperatures.
This polarization can be transferred to the deuterium nuclei by an
Overhauser type effect through dynamic nuclear polarization, where a
single microwave transition is saturated.    
The relatively short spin lattice relaxation
time of the electrons coupled with their high photon energy provides a
large effective heat capacity from which to cool the deuterons. In
existing experiments, diamagnetic crystal polarizations of greater
than 90\% were achieved by such means~\cite{abragam:qc1982a}.  An
experiment more closely related to this proposal is De Boer's
polarization of a deuterated sample of ethanediol (a sample containing
$^{13}C$, $^1H$ and $^2D$ nuclei).  The polarization for $^1H$ and
$^2D$ reached 80\% and 30\%, respectively~\cite{deboer:qc1974a}.  The
cooling of the deuterium bath to sub-mK spin temperatures is a preparation
step that is done before starting the computation.  Following this the
electrons will no longer take part in any aspect of the computation
and the deuterium will act as a polarized bath of nearly infinite heat
capacity that the qubits of the QP can be brought into contact at
will.  Connection to the polarized deuterium bath will be accomplished
through Hartmann-Hahn cross polarization transfer~\cite{hartman:qc1962a}, where
two nuclear species are put into an interaction frame in which their
apparent Zeeman energies are matched.

Since the QCs are only very dilute in the deuterium bath, the heat
capacity of the bath will be much greater than that of the QC meaning
that the final polarization of the QC and bath (following
cross-polarization) will be very nearly the initial polarization of
the bath and the bath may be in contact repeatedly without a need to
refresh it from the electrons.  Of course the suggestion that the bath
may be reused implies that as we decouple it 
we must do so in such a way as to
preserve its polarization.

Finally, although the QC has been prepared in a state that is quite
close to a pure state, we will still need to rely on pseudo-pure or
state purification schemes~\cite{schulman:qc1998a}, perhaps in
conjunction with the resettable register, to achieve a reliable pure
state.

\subsubsection{Gates}

The internal Hamiltonian in conjunction with the coupling to RF fields
provide all of the components necessary to create, in principle, any
gate.  Under the assumption that the resonance frequencies of the
spins are sufficiently far apart, single qubit gates can be
implemented by selective RF pulses, while two-qubit gates are
implemented by delays which exploit the connected coupling network of
the internal Hamiltonian. Appropriate refocusing schemes are required
to eliminate unwanted contributions from the internal Hamiltonian. In
particular, there must be a means for suppressing the internal
Hamiltonian completely.  If we can suspend the evolution of the
internal motions, then we can always add back in any parts we
desire. In earlier work we and others have devised time-suspension
sequences that suppress the internal Hamiltonian to any desired
precision~\cite{cory:qc1990a} and have shown how to add back desired
pieces of it~\cite{munowitz:qc1988a}. The complications introduced
here are that the gates should never interfere with the procedure for
$^2D$ decoupling of the bath from the QP.  This requires synchronizing
the various multiple-pulse cycles and designing them from the start to
work together.

\subsubsection{Noise}

Decoherence times of dielectric crystals below their Debye temperature
are notoriously long in NMR since there is no spontaneous emission
and the phonon spectrum is very sparse.  The only relaxation pathway is
via defects, and thus will be very sample dependent. The relaxation rate
$1/T_1$ goes as $(H_e/H_0)(1/T_{1e}) (1-p^2)$ where
$H_e$ is the field due to the electrons, $H_0$ is the applied magnetic
field, $T_{1e}$ is the electron relaxation time and $p$ the electron
polarization.  So in this approximation, as the electrons become
polarized, the nuclear relaxation rate vanishes~\cite{gunter:qc1967a}.

In a system of dilute molecules with the spin transport through the
lattice turned off through multiple-pulse decoupling, we expect that
the decoherence times will be even longer than those typically
observed for concentrated samples of abundant
spins~\cite{mavlonazarov:qc1992a}.  Times on the order of hours are
common (malonic acid crystals have a $T_1$ of order one
hour) and for clean cubic systems there are examples
of relaxation times being so long as to be unmeasurable.  Even given relatively slow clock speeds  in the ms, it
is clear that spin lattice relaxation will not significantly limit the
computational complexity of the proposed quantum computer.

Of course there is the related issue of the spin-spin relaxation time.
Any difference between the spin-spin and the spin-lattice relaxation
times are due to spin-spin interactions within the QP and these
interactions must be precisely controlled to have a functional quantum
computer.  This puts a premium on our ability to apply average
Hamiltonian theory, with multiple pulse cycles to take full control of
the spin system, but this is exactly where NMR spectroscopy excels.

\subsubsection{Measurement}

The measurement process does not significantly differ from that used
in liquid state NMR.  The sample is still an ensemble of processors,
and though the polarization is higher and the sample much more dilute,
there will still be a sufficient number of spins to have good
statistics.

\section{Future generations QIP based on engineered control of spin systems}

Although we hope to reach quite complex Hilbert spaces with the solid
state NMR approach to QIP, this still does not yield a scalable
quantum computer.  The number of qubits is limited by the addressing
scheme, the gates are far too slow, and the cost of additional qubits
too high.  While the progression towards more complex QIPs is stepwise, 
it is useful to have a trajectory in
mind for a path to a scalable quantum computer.

Keeping to our model of a QP as a quantum simulation
run on a physical system, there are two general approaches to implement
the required dynamics:
\begin{itemize}
\item[1)] Use the internal Hamiltonian and limited external
control and move to an appropriate interaction frame to enable qubit-specific operations. This approach is limited as we can only scale down part of the Hamiltonian. 
\item[2)] Modify the local Hamiltonian by near field interactions
that are qubit-selective based on their spatial position and extent.
\end{itemize}

It appears that the second (or engineered) approach will be the
eventual direction, but control that relies on the internal
Hamiltonian is what can be done with high fidelity today.  Although
there have been some steps taken towards engineered
systems~\cite{awschalom:qc2000a,kane:qc1997a,yablonovitch:qc1999a}, 
little is known about how to
manufacture such devices, and even less about how to control them.

As the field progresses towards more complex QIP it appears that there
are three regimes of complexity/decoherence:
\begin{itemize}
\item[1)] Small scale QPs: Fully
coherent control over a small quantum system, decoherence is not a
significant problem and the dynamics could be simulated on a work
station---this is where we are today.
\item[2)]  Medium scale QPs:
Complex and high fidelity coherent control of a modest quantum system. 
Decoherence is present but we can demonstrate accessibility
of the full Hilbert space.  The dynamics are no longer easily
simulated on classical computers---this is where solid
state NMR QIP is headed.
\item[3)] Large scale QPs: Local near field control of QPs' components
making general transformations of many qubits much more accessible.
The effects of decoherence on these systems is much more significant
requiring sophisticated error-control schemes.
\end{itemize}

The proposed solid state NMR device is intended to be medium scale,
allowing exploration of problems not classically accessible
while avoiding the decoherence problems of the more scalable
near-field devices.

\section{Conclusion}

By enabling reasonably reliable manipulations of quantum information,
NMR offers a useful test bed for QIP.  The elements essential
\cite{divincenzo:qc2000a} for quantum information have been
demonstrated.  It is also moving theoretical successes from the desk
to the laboratory benchtop and helps in understanding the difficulties
of harnessing the power of quantum computation.  NMR techniques have
been developed in the spectroscopy community to deal with a wide range
of coherent errors and although the final technology for quantum
computation might not resemble what we can imagine today, the set of
techniques being developed are needed to make progress.

Another important contribution from NMR has been to question some of
the naive explanations of the power of
quantum computation such as the need for pure states and entanglement. As a result, it has been realized that
QIP extends well beyond quantum computing or communication,
and the relative computational power of devices which
do not fit the standard model is starting to be explored.

Despite the limitations of liquid state NMR, it is likely to be the
first technology to control ten qubits.  Many of these
limitations will be overcome by a second generation of NMR based
devices which has the potential to exceed the classical predictable
realm.  Our solid state proposal, by starting with high polarization
and having a resettable register, allows a better understanding of
practical quantum error correction.  It is our hope that it will also
allow a better understanding of quantum simulation and decoherence.  As QIP devices
access increasing numbers of qubits, it becomes increasingly difficult
to simulate them using classical computers (and is impossible at
around 30-40 qubits).  This should be seen as an opportunity to
investigate a new territory inaccessible to classical simulations.
Systematic ways to deal with the complexity of these systems will have
to be developed such as benchmarking and trouble shooting
imperfections of physical devices.  On the other hand the number of
qubits available in solid state NMR is still small compared to what is
needed for implementing the gamut of fault tolerant techniques.  We
will thus have to develop methods to investigate extremely large
Hilbert spaces and verify the quality of the implemented operations.
The work in NMR will contribute to these developments.

The recent discovery that quantum mechanics allows more powerful
manipulations of information than its classical counterpart has the
potential to revolutionize information processing.  Although the
accuracy threshold theorem shows that in principle, the limitations of
quantum systems due to noise and decoherence can be overcome, this
does not mean that the road to scalable QIP will be without difficulties. 
The required
accuracies will be hard to achieve in practice, and at present
we can only explore small scale quantum information.

It is of course possible that we will not be able to reach the
accuracy threshold nor ever implement algorithmically interesting
quantum computations.  However, as more than 100 spin coherences have
been manipulated, controlled and observed in solid state NMR, the main
concern is our ability to maintain information in effectively
distinguishable qubits. This is both a question of complexity and of
stability.  So while we are still struggling with the new quantum
technology and its control, we are approaching the situation where we
can ask whether there is any new physics which prohibits scalable
manipulations of quantum information. As a result, the significance to
practical and fundamental physics promises to be profound.

\noindent {\bf Acknowledgements.} This work was suppported by the U.S. Army
Research Office, DARPA and the National Security Agency.   DGC thanks the Center
for Non-linear Systems at Los Alamos National Laboratory for its hospitality.

\bibliographystyle{plain}

\end{document}